# Creation of electrical knots and observation of DNA topology

Tian Chen[1*], Xingen Zheng[1*], Qingsong Pei, Deyuan Zou, Houjun Sun, and Xiangdong Zhang[1+]

[1] Key Laboratory of advanced optoelectronic quantum architecture and measurements of Ministry of Education, Beijing Key Laboratory of Nanophotonics & Ultrafine Optoelectronic Systems, School of Physics, Beijing Institute of Technology, 100081, Beijing, China

*These authors contributed equally to this work. +Author to whom any correspondence should be addressed. E-mail: zhangxd@bit.edu.cn

## Abstract

Knots are fascinating topological structures that have been observed in various contexts, ranging from micro-worlds to macro-systems, and are conjectured to play a fundamental role in their respective fields. In order to characterize their physical properties, some topological invariants have been introduced, such as unknotting number, bridge number, Jones Polynomial and so on. While these invariants have been proven to theoretically describe the topological properties of knots, they have remained unexplored experimentally because of the difficulty associated with control. Herein, we report the creation of isolated electrical knots based on discrete distributions of impedances in electric circuits and observation of the unknotting number for the first time. Furthermore, DNA structure transitions under the action of enzymes were studied experimentally using electrical circuits, and the topological equivalence of DNA double strands was demonstrated. As the first experiment on the creation of electrical knots in real space, our work opens up the exciting possibility of exploring topological properties of DNA and some molecular strands using electric circuits.

# I. INTRODUCTION

The exploration of knot physics has become one of the most fascinating frontiers in recent years, due to its complex topology that plays an important role in physical and life sciences [1-5]. At present, various knots have been constructed based on different physical implementations. According to the characteristics of the matter that forms the structure, the knots can be divided into two types. One is called a "continuous" knot, which is formed by continuously distributed substances, such as knots in liquid crystal [6, 7], fluid [8-12], elastic media [13], momentum space [14], and light and acoustic systems [15-23]. The other is called the "discrete" knot, which is formed by discrete lattices, such as knots in DNA and other molecular strands [24-32].

"Discrete" knots are usually found in the microscopic world, particularly on the cellular level. Different molecular knots can be synthesized through the accurate control of chemical reactions [1, 33-38]. After acting on duplex cyclic DNA molecules with direct repeats, enzymes called topoisomerases are able to generate different nontrivial DNA knots. Topoisomerases have shown to be tendentious in knotting or unknotting DNA molecules, which have completely different functions in living systems. For instance, type II topoisomerase manifests a preference to unknot DNA molecules. Therefore, understanding the possible topologies of DNA molecules is helpful to explore life mechanisms. In recent years, many theoretical investigations, including lattice-based simulations, equilateral chain model simulations, grid diagrams and so on, have proven to generate the topology of different DNA knots [35-38]. However, the direct experimental evidence of DNA topology, especially the topological equivalences between two different DNA structures, is still lacking.

In this work, we experimentally constructed the "discrete" knots based on circuits [39-48], which differ from other realizations of "continuous" knots for macroscopic objects. The advantage of constructing such an experimental platform is that it can be used to study various phenomena corresponding to DNA topology. For example, through the change of impendence in the electric circuits, we could for the first time have been able to render one important topological invariant in knot theory experimentally observable: namely, the unknotting number. The topological change of DNA molecules under the action of enzyme Tn3 resolvase was also observed. In addition, the topological equivalence between two different structures of DNA molecules was experimentally demonstrated using the circuit platforms.

## II. ELECTRICAL REALIZATION OF KNOTS AND OBSERVATION OF UNKNOTTING NUMBER

To implement electric knots in an electric circuit, the following steps were performed. At first, we established a connection between the knot theory and the corresponding construction in the lattice. The coupling strengths between sites in the lattice were designed, and the sites in the lattice occupied by the localized eigenstates comprised the knot structure. Then, we designed the electric circuit and correlated the Laplacian describing the circuit to the Hamiltonian in the lattice by choosing the appropriate electric capacitors and grounding elements. Finally, the distributions of impedance in the circuit were measured, where nodes possessing large impedances correspond to the sites occupied by localized eigenstates in the lattice and form the knot. The detailed construction method is described in Appendix A. Following such a method, various electrical knots were created. Figs. 1a and 1b display the designs of three-layer electric circuits, which were integrated with a trefoil knot and unknot structures, respectively. The red spheres represent the nodes possessing large impedances (>1 k$\Omega$), and other nodes have an impedance smaller than 0.1 k$\Omega$. The purple tubes indicate the connections among the red spheres through the small electric capacitor.

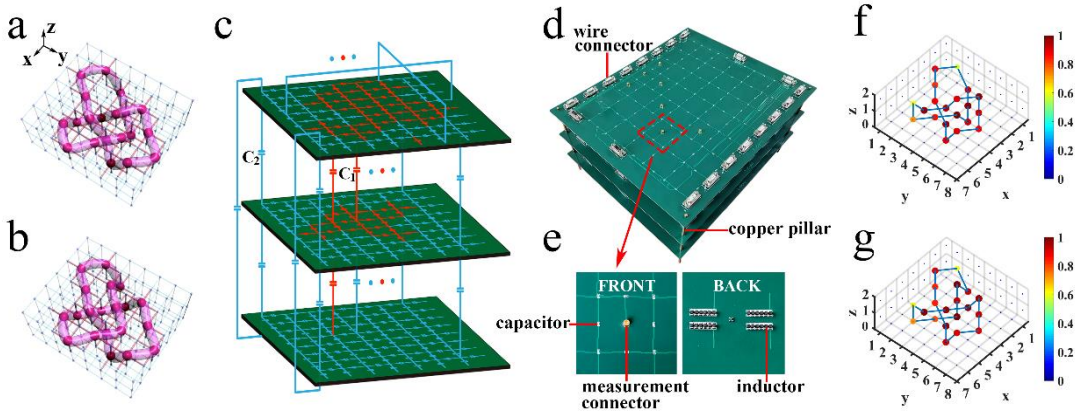

**Fig. 1. Electric realizations of knots. (a) The construction of the trefoil knot in the electric circuit. The red cylinders represent the small electric capacitors $C_1$ = 100pF, and the blue cylinders denote the large electric capacitors $C_2$ = 10nF. Large red spheres indicate that the nodes possessing the impedance larger than 1 k$\Omega$, and other nodes have impedances smaller than 0.1 k$\Omega$. The connection composed by red spheres is exactly the trefoil knot. (b) The**

**construction of the unknot in the electric circuit. (c) Details of the electric design for the trefoil knot in (a). (d) The experiment setup for the trefoil knot. It contains three printed circuit board (PCB) layers. (e) We zoom in one plaquette containing four nodes at the front of one layer. Each node in the circuit is linked with the capacitor. The measurement connector is provided to measure the impedance through the impedance analyzer. The grounding inductors are provided at the back of each layer. (f) and (g) The distributions of impedance obtained from the experiments for the trefoil knot (panel a) and the unknot (panel b). The value of impedance at each node has been normalized. We connect the nodes with impedance larger than $1\,\text{k}\Omega$ by one solid line.**

In order to observe such a phenomenon experimentally, we designed the corresponding circuits (the circuit corresponding to Fig. 1a is shown in Fig. 1c). For the convenience of experiment, the total cube was cut into three layers, which were then positioned on three printed circuit boards (PCBs). Capacitors were connected to every node on adjacent layers. A cyclic boundary condition was applied to avoid the edge effect. The experimental setup of the fabricated circuit is shown in Fig. 1d. In one PCB layer, the capacitors are arranged on the front and grounding inductors on the back. The enlarged image of the front and back of a plaquette is given in Fig. 1e. The fabricated sample has exactly the same construction as that shown in Fig. 1a. The measurement results of impedances corresponding to the cases in Figs. 1a and 1b are presented in Figs. 1f and 1g, respectively. Experimental details are given in Appendix B. The dots in Figs. 1f and 1g represent the nodes with impedance larger than $1\,\text{k}\Omega$. Accordingly, the experimental results were found to agree well with the results presented in Figs. 1a and 1b.

Because circuit networks possess remarkable advantages of being tunable and reconfigurable, many interesting problems associated with the knot theory using electric knots can be explored. In mathematics, some invariants are usually introduced to describe the topological properties of knots, such as unknotting number, bridge number, Jones Polynomial, and Alexander Polynomial [1,4,5]. Although these concepts and theories have existed for around 100 years, they have never been proven by experiments. For the first time, this work provides the observation of these topological invariants in the designed circuit platforms. Particularly, one invariant, the unknotting number, is defined as the least times in the change of crossing to make the knot become an unknot. Thus, the key to observe the unknotting number is to realize the change of crossing in the electric circuit.

Figures 2a-e illustrate the change of one crossing in geometry, whereby such type of evolution process can be shown in the designed circuits. Figures 2f-j present the corresponding circuits, and construction details are provided in Appendix C. The red spheres in Figs. 2f to 2j denote the nodes with impedances larger than 1 k$\Omega$, while other nodes have impedances smaller than 0.1 k$\Omega$. By modulating the capacitors between nodes in a sequence, we can obtain the change of impedance distribution continuously, meaning that the change of crossing can be well demonstrated in the designed circuits. It should be noted that the interference appears in the hopping rates of two adjacent nodes (the broken parts of tubes) when the purple tube is very close to the orange tube (Figs. 2g and 2i), which leads to smaller impedances at these nodes simultaneously. However, such a phenomenon does not have any influence on counting the times of the crossings change in the circuit.

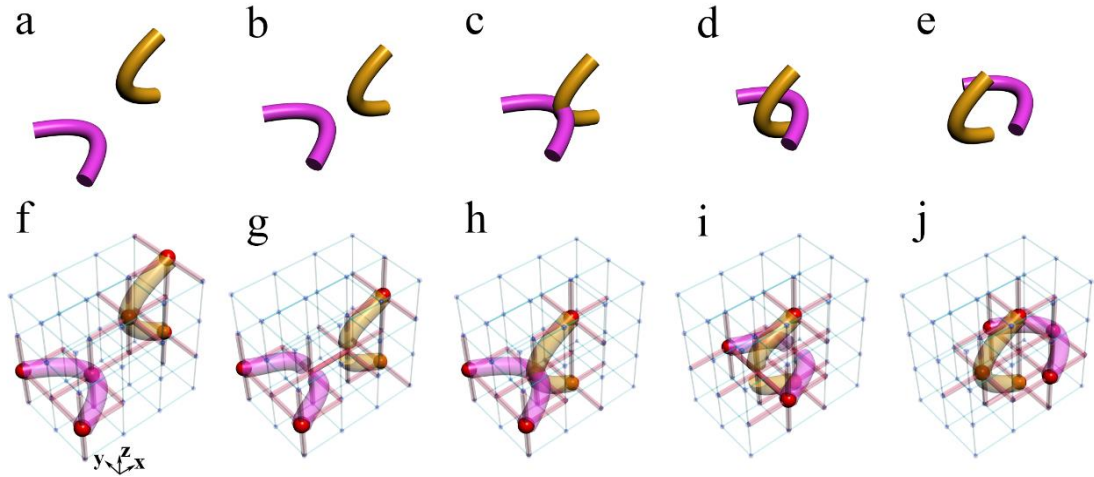

**Fig. 2. The change of one crossing in geometry and electric realizations. (a)-(e) We provide the process to change the crossing between the purple and the orange tube once. (f)-(j) Circuit designs to illustrate such changing process. In each panel, red (blue) cylinders between every two nodes represent the electric capacitors $C_1$=100pF ($C_2$=10nF). Red spheres denote the nodes with large impedances. We use purple and orange tubes to connect these nodes with large impedances.**

We further studied the change of crossing in the knot structures, considering the transition from the electrical trefoil knot in Fig. 1a to unknot in Fig. 1b. In the transition process, the change of crossing occurs only once, which was observed in the designed circuit by continuously modulating the capacitors and associated grounding elements. The detailed discussions and experimental results are presented in Appendix D. This observation indicates that the unknotting number of the trefoil knot is 1, according to the definition [1]. A similar case was also seen for the figure-8 knot. In contrast, for the transition from the $8_3$ knot to unknot, two crossing changes were observed, thus, the unknotting number for the $8_3$ knot is 2. A detailed discussion is addressed in Appendix D.

Similarly, other invariants such as the bridge number, Jones Polynomial and Alexander Polynomial, can also be observed experimentally by designing the corresponding circuits. An important feature of electric knots is their discrete distributions of impedances, which enables us to study the topological properties of molecular strands. For example, the phase transition of DNA structures and topological equivalence problems of DNA double strands can be discussed by designing circuit platforms.

## III. OBSERVATION OF THE TOPOLOGICAL PHASE TRANSITION IN DNA STRUCTURES USING ELECTRICAL CIRCUITS

It is known that the duplex cyclic DNA molecule can show the structures of knots and links under the action of topoisomerases. In living systems, knotting and unknotting DNA molecules have totally different functions, whereby different knots and links of DNA molecules can appear when enzyme Tn3 resolvase is applied with direct repeats [1,2,33,34]. Before the action of Tn3 resolvase, the DNA molecule can be viewed as two tangle structures, as shown in Fig. 3a, where one substrate tangle S is not affected by Tn3 resolvase and the other site tangle T is influenced by Tn3 resolvase. In this case, the DNA molecule is unknotted. After one action of Tn3 resolvase, the site tangle T is replaced by the recombination tangle R, and the geometric structure of DNA molecule changes to that of a one hopf link, as shown in Fig. 3d.

The corresponding circuits can be constructed to display the above phenomenon very well. Figs. 3b and 3e display the designed circuits corresponding to the structures in Figs. 3a and 3d, respectively. In the circuits, different electric capacitors are chosen to connect the two nearest neighboring nodes. In the figure, the structural functions of DNA molecules are integrated into the design of circuits, where red spheres denote the nodes with impedances larger than $1\,\mathrm{k\Omega}$, and other nodes have impedances smaller than $0.1\,\mathrm{k\Omega}$. These red spheres in Figs. 3b and 3e exactly constitute the connection representing the structures shown in Figs. 3a and 3d, respectively. The designed circuits were fabricated successfully, and the measured impedances from experiments are shown in Figs. 3c and 3f. Details of circuit constructions and experiments are given in Appendix E. Comparing these measured results with the corresponding structures in Figs. 3a and 3d reveals good agreements.

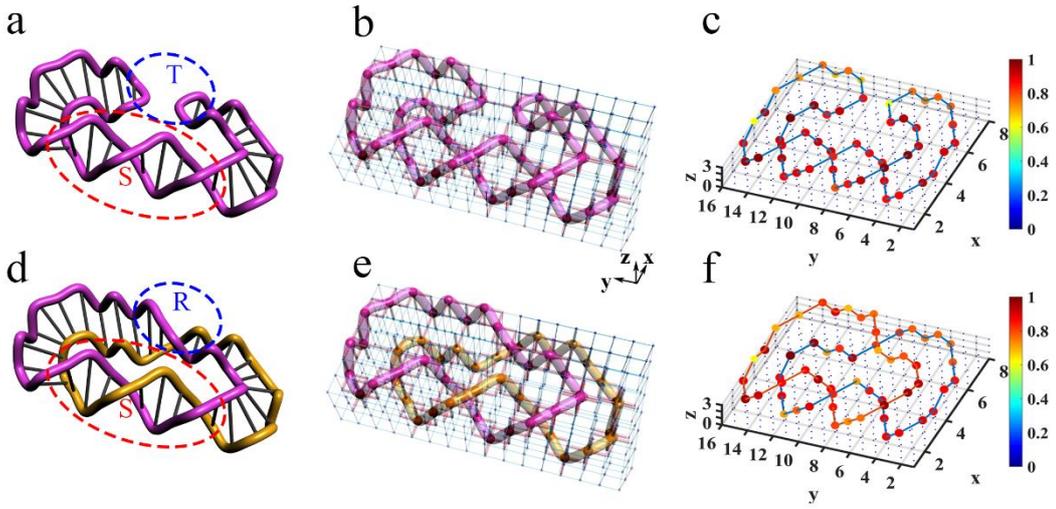

**Fig. 3. Electrical realizations of different DNA molecular structures before and after the action of topoisomerase. (a) and (d), the structures of DNA before and after the action of the enzyme Tn3 resolvase. In (a), the structure of DNA is composed by one substrate tangle S and the other site tangle T. In (d), after the action of Tn3 resolvase, the site tangle T is replaced by the recombination tangle R. (b) and (e), the electric realizations of different DNA structures before and after the action of Tn3 resolvase. In each panel, red (blue) cylinders between every two nodes represent the electric capacitors $C_1$=100pF ($C_2$=10nF). Red spheres denote the nodes with large impedances. We connect these nodes in purple and orange tubes. (c) and (f), the distributions of impedances in experimentally electric circuits. All the values of impedance are normalized to the largest impedance in each panel. We use the solid lines to connect the nodes with large impedance.**

The above discussions present only one phase transition of DNA molecules from the unknotted structure to the hopf link under the action of topoisomerases. In fact, multiple phase transitions can occur when the topoisomerases react repeatedly. For example, the structure of DNA molecules in Fig. 3d transitioned into the figure-8 knot structure when enzyme Tn3 resolvase was continuously applied. Such multiple phase transitions can also be well demonstrated in designing circuits by modulating the capacitors and associated grounding elements (see Appendix E). That is to say, all kinds of phase transitions of DNA molecules can be well observed by designing the circuits, further demonstrating the powerful ability to study related problems in this way.

## IV. DEMONSTRATION OF TOPOLOGICAL EQUIVALENCE OF DNA DOUBLE STRANDS IN THE ELECTRIC CIRCUIT

DNA double strands play a dominant role in various biological functions that are necessary for life, such as replication, transcription and recombination [1,2]. Cyclic duplex DNA exhibits different geometric structures under different ambient environments. Figure 4a shows the cyclic duplex DNA structure in the relaxed state, which becomes an intermediate state (Fig.1d) with the change of ambient environment, and then transitions into an unrelaxed state (Fig. 4g). In the unrelaxed state, the DNA structure is more tightly twisted. To describe the different geometric structures of cyclic duplex DNA, the twist ($Tw$) is defined as how much the DNA structure twists around its axis, and the writhe ($Wr$) refers to how much the axis of the DNA structure is contorted in space [1,49,50]. When the cyclic duplex DNA structure is in the relaxed state (Fig. 4a), it does not twist around its axis; instead, its axis is contorted once in space and, thus, the $Tw = 0$ and $Wr = 1$. In Fig. 4g, the axis of the DNA structure lies flat in the plane, but the structure twists around its axis once, and thereby, $Tw = 1$ and $Wr = 0$.

The topology of the DNA geometric configuration determines the functions of living mechanisms. The linking number $Lk = Tw + Wr$ is usually introduced to characterize the topological properties of different geometric configurations. It has been proven that, although the geometric configuration is different, the linking number is the same, and the function of DNA is the same. This is called the topological equivalence. For example, the two geometric configurations shown in Fig. 4a and 4g seem different, yet their linking numbers are the same.

The demonstration of topological equivalence between two different geometric configurations is very meaningful, but is very difficult to experimentally demonstrate in living cells. Using the designed circuit platforms, we were able to observe the phenomena very well. By integrating the cyclic duplex DNA structure function in the relaxed state into the circuit design, we constructed a circuit that displays the DNA configuration (Fig. 4b). Similar to the above figures, red spheres denote the nodes with impedances larger than $1\,k\Omega$, and other nodes have impedances smaller than $0.1\,k\Omega$. It is seen clearly that the red spheres in Fig. 4b exactly constitute the connection representing the structure shown in Fig. 4a.

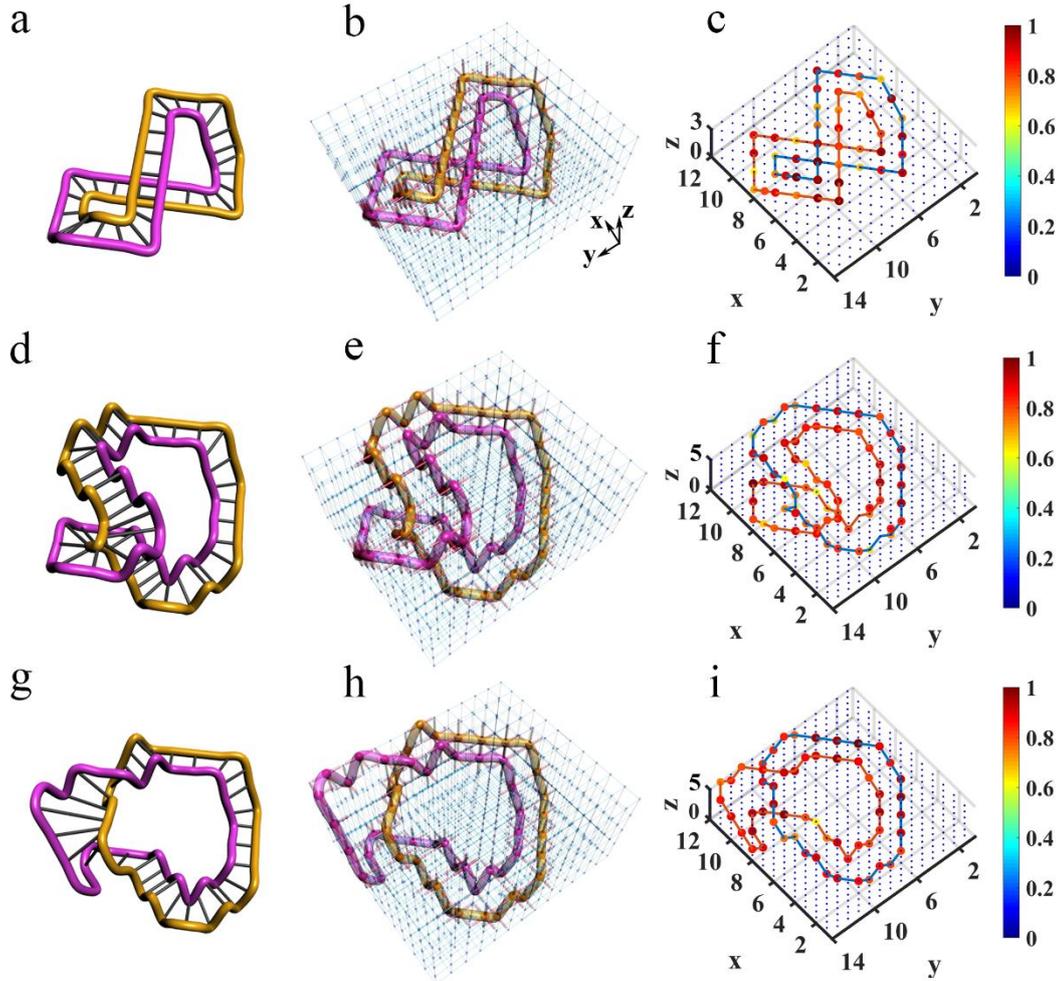

Fig. 4. Electric demonstration of topological equivalence between different DNA structures. (a) The simplest DNA structure with the twist $\text{Tw}=0$ and the writhe $\text{Wr}=1$. (b) The electric design of (a). (c) The experimental distribution of impedance in the electric circuit for (b). (d) The DNA structure in the middle of the change process. (e) The electric design of (d). (f) The experimental distribution of impedance for (e). (g) The DNA structure with the twist $\text{Tw}=1$ and the writhe $\text{Wr}=0$. (h) The electric design of (g). (i) The experimental distribution of impedance for (h). In panel (b), (e) and (h), red (blue) cylinders between every

**two nodes represent the electric capacitors $C_1$=100pF ($C_2$=10nF). Red spheres denote the nodes with large impedances. In panel (c), (f) and (i), all the values of impedances are normalized to the largest one, and we use the solid lines to connect the nodes with large impedances.**

By tuning the capacitors and associated grounding elements, the corresponding circuits for the intermediate state (Fig. 4d) and unrelaxed state (Fig. 4g) were also obtained, which are shown in Figs. 4e and 4h, respectively. The fabricated details and measured process for these designed circuits are described in Appendix F. The measured results of impedances are given in Figs. 4c, 4f, and 4i, which demonstrated good correspondence to the cases in Figs. 4a, 4d, and 4g. While the linking number $Lk = 1$ was proven for the two structures in Figs. 4c and 4i, it is not enough to confirm their topological equivalence.

Therefore, to demonstrate the topological equivalence between the structures, a continuous change from one structure to another needed to occur. That is, there is no change of crossing in the process. In fact, in the designed circuits, it is convenient to study the continuous change between two structures by modulating the electric capacitors and related grounding inductors step-by-step. Details of the change process for such a case are given in Appendix F. Results reveal that there is no change of crossing occurred in the process of continuous change between the two structures in Fig.4c and 4i, thereby verifying their topological equivalence.

## V. CONCLUSION

This work theoretically and experimentally proposes schemes to create isolated electrical knots based on discrete distributions of impedances in electric circuits. Using the designed circuit platforms, we experimentally observed unknotting numbers for the first time, and the corresponding DNA structure transitions under the action of enzymes using electrical circuits. The topological equivalence of DNA double strands was further demonstrated experimentally. Our results provide convincing evidence that the electric circuits offer reliable platforms to study the various topological properties of knots and links, especially microscopic "discrete" knots in the molecular strands. Moreover, such circuit platforms revealed new phenomena in this work and also provide a basis for future exploration of unsolvable problems.

**ACKNOWLEDGMENTS**

This work was supported by the National key R & D Program of China under Grant No. 2017YFA0303800 and the National Natural Science Foundation of China (No.91850205 and No.61421001).

**Appendix A: The general scheme to create electrical knots**

Here, we describe how to construct knots and links in electric circuits. We start from the standard procedure of obtaining the knots and links, and then provide the design on the circuit.

1. Construction of standard knots

According to the theory [1,2], since the abstract group definitions of braids have remarkable geometric and topological interpretation, the knots and links can be obtained from the braids. The braided function is expressed with a formal variable $u$ and has $N$ distinct zeros at positions parameterized by the height $h$ in Eq. (A1) as

$$p_h(u) = \prod_{j=0}^{N-1}(u - (\cos h_j + i\sin \beta h_j)). \tag{A1}$$

Since the knots or links are composed by closed braids, the braid function needs to be periodic in $h$. It is convenient that the range of $h$ is taken from 0 to $2\pi$, so after $h$ changed by $2\pi$, the zeroes in the complex $u$ plane are in the same positions as where they started. Moreover, the term $e^{ih}$ can be substituted by a second formal variable $v$. After this process, we can obtain the expression for different knots and links with these two formal variables $u$ and $v$. E.g., $p_h(u)=u^2 - v^2 =0$ represents the standard hopf link and $p_h(u)=u^2 - v^3 =0$ denotes the standard trefoil knot.

The formal variables in the expression above can be mapped to the $R^3$ space through the Milnor map. We use Cartesian coordinates $x, y,$ and $z$ to express the formal variables $u$ and $v$ in the following equations:

$$u = \frac{x^2+y^2+z^2-1+2iz}{x^2+y^2+z^2+1} = \frac{R^2+z^2-1+2iz}{R^2+z^2+1}$$
$$v = \frac{2(x+iy)}{x^2+y^2+z^2+1} = \frac{2Re^{i\phi}}{R^2+z^2+1}$$
(A2)

Given in both Cartesian and cylindrical coordinates, $R^2 = x^2 + y^2$ and $Re^{i\phi} = x + iy$, the expression for the standard trefoil knot is written as

$$f_{tre}(\vec{r}) = \left[(R^4-1)(R^2-1) - 8R^3 e^{i3\phi}\right] + 4iz(R^4-1)$$
$$+ z^2(3R^4 - 6R^2 - 5) + 8iz^3 R^2 + z^4(3R^2 - 5) + 4iz^5 + z^6 = 0$$
(A3)

In Fig. 5a, we provide the standard trefoil knot in the $R^3$ space.

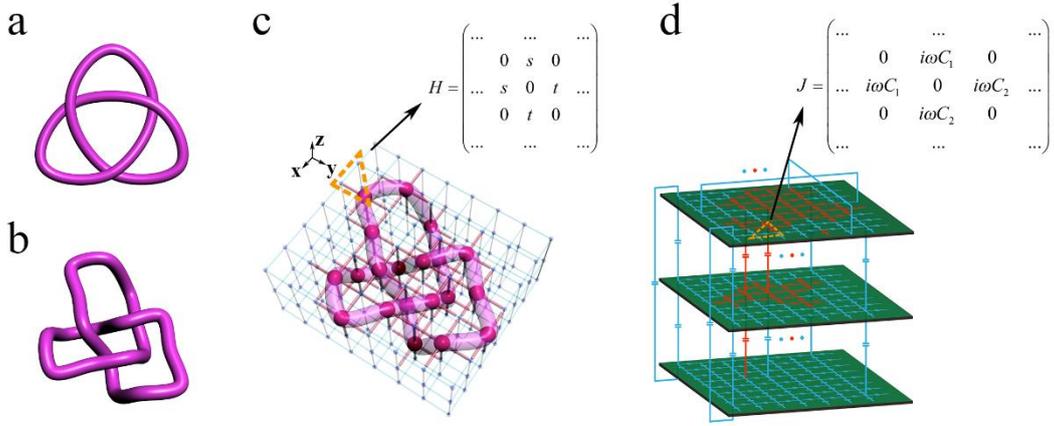

**Fig. 5. Details of electrical construction of trefoil knot in the circuit.** (a) The standard trefoil knot. (b) The deformation of standard trefoil knot. (c) Construction of the trefoil knot (b) in the lattices. Red (blue) cylinders represent the couplings strengths $t = 0.01$ ($s = 1$). The Hamiltonian for three sites in the lattice is presented. (d) Realization of the trefoil knot (b) in the electric circuit. The value for capacitors in red (blue) is $C_1 = 100$pF ($C_2 = 10$nF). The Laplacian for three nodes in the circuit is presented explicitly.

The mathematic theory tells us that the knots and links can be deformed freely in space, but not allowed to be cut off or glued. After such a deformation process, the geometric structure representing the knot (link) can be changed to another structure. Although these two structures seem different at the first sight, they are actually the same knots (links). It is often called these two knots (links) are isotopic. It is hard to find an ambient isotopy function in general, the researchers often project the knot into two-dimension and manipulate this projection in three formal ways, i.e., the Reidemeister moves. These manipulations are equivalent to an ambient isotopy in 3D. In Fig. 5b, we provide one structure obtained from the deformation of the structure in Fig. 5a. These two are isotopic since they can be deformed into each other without being cut off or glued. The expression for the deformed trefoil knot in Fig. 5b is shown as

$$f_{deformed-tre} =$$
$$-1 + a_1 x + a_2 y + a_3 z + a_4 x^2 + a_5 xy + a_6 xz + a_7 y^2 + a_8 yz + a_9 z^2 + a_{10} x^3$$
$$+ a_{11} x^2 y + a_{12} x^2 z + a_{13} xy^2 + a_{14} xz^2 + a_{15} xyz + a_{16} y^3 + a_{17} y^2 z + a_{18} yz^2 + a_{19} z^3 + a_{20} x^4$$
$$+ a_{21} x^3 y + a_{22} x^3 z + a_{23} x^2 y^2 + a_{24} x^2 yz + a_{25} x^2 z^2 + a_{26} xy^3 + a_{27} xy^2 z + a_{28} xyz^2 + a_{29} xz^3 + a_{30} y^4 \quad (A4)$$
$$+ a_{31} y^3 z + a_{32} y^2 z^2 + a_{33} yz^3 + a_{34} z^4 + a_{35} x^6 + a_{36} x^4 y^2 + a_{37} x^2 y^4 + a_{38} y^6 + a_{39} zx^4 + a_{40} zx^2 y^2$$
$$+ a_{41} zy^4 + a_{42} z^2 x^4 + a_{43} x^2 y^2 z^2 + a_{44} z^2 y^4 + a_{45} z^3 x^2 + a_{46} z^3 y^2 + a_{47} z^4 x^2 + a_{48} z^4 y^2 + a_{49} z^5$$
$$+ a_{50} z^6 + a_{51} x^5 + a_{52} y^5$$

The deformed trefoil knot can be composed of two fitting curves when we set $|f_{deformed-tre}| \leq 5*10^{-9}$. The ranges of coordinates are, $x \in [2,7]$ and $y \in [2,8]$, $z \in [1,2]$. The values from $a_1$ to $a_{52}$ for these two fitting curves are listed below,

**Table 1.** The coefficients $a_1$ to $a_{52}$ in the $f_{deformed-tre}$. The values outside and inside [*] are the coefficients of the first and second fitting curves, respectively.

| | |
|---|---|
| $a_1$ = 0.2353-0.0260i [-0.0332+0.0187i] | $a_2$ =0.0428+0.0102i [0.2930+0.0014i] |
| $a_3$ =1.2146-0.0230i [0.5328+0.0342i] | $a_4$ =0.0120+0.0014i [0.0226+0.0030i] |
| $a_5$ =-0.0080+0.0050i [-0.0091+0.0021i] | $a_6$ =-0.2362+0.0198i [-0.0899-0.0134i] |
| $a_7$ =-0.0031+0.0032i [0.0015-0.0065i] | $a_8$ =-0.0918-0.0084i [-0.0629+0.0063i] |
| $a_9$ =-0.0973+0.0038i [0.1002-0.0328i] | $a_{10}$ =(4.14-6.47i)*10$^{-4}$ [(1.39+2.17i)*10$^{-4}$] |
| $a_{11}$ =-0.0008-0.0011i [0.0001-0.0012i] | $a_{12}$ =-0.0133-0.0018i [0.0033-0.0012i] |
| $a_{13}$ =(6.07+6.39i)*10$^{-4}$ [-0.0014+0.0007i] | $a_{14}$ =-0.0009+0.0017i [-0.0061+0.0021i] |
| $a_{15}$ =-0.0008-0.0053i [-0.0077-0.0027i] | $a_{16}$ =-0.0002+0.0011i [-0.0025+0.0007i] |
| $a_{17}$ =0.0067-0.0053i [-0.0102-0.0009i] | $a_{18}$ =0.0063+0.0090i [-0.0179+0.0002i] |
| $a_{19}$ =-0.0961-0.0017i [0.0048+0.0009i] | $a_{20}$ =(-5.03+2.24i)*10$^{-4}$ [2.40*10$^{-5}$-2.21*10$^{-4}$i] |
| $a_{21}$ =1.99*10$^{-7}$+9.17*10$^{-11}$i [2.04*10$^{-4}$-6.26*10$^{-5}$i] | $a_{22}$ =(-2.00+3.23i)*10$^{-4}$ [2.82*10$^{-5}$-1.02*10$^{-4}$i] |
| $a_{23}$ =(-1.49-2.91i)*10$^{-4}$ [(4.98-2.15i)*10$^{-4}$] | $a_{24}$ =(4.17+5.27i)*10$^{-4}$ [(5.68-1.36i)*10$^{-4}$] |
| $a_{25}$ =0.0075-0.0009i [0.0028+0.0004i] | $a_{26}$ =1.43*10$^{-8}$-1.34*10$^{-10}$i [-2.04*10$^{-4}$+6.26*10$^{-5}$i] |
| $a_{27}$ =(-3.04-3.20i)*10$^{-4}$ [(8.81-3.23i)*10$^{-4}$] | $a_{28}$ =0.0024+0.0014i [0.0027+0.0010i] |
| $a_{29}$ =0.0301-0.0025i [(-9.02-9.53i)*10$^{-4}$] | $a_{30}$ =4.80*10$^{-5}$-1.67*10$^{-4}$i [(-7.31-1.66i)*10$^{-5}$] |
| $a_{31}$ =8.69*10$^{-5}$-5.31*10$^{-4}$i [0.0018-0.0001i] | $a_{32}$ =0.0002+0.0015i [-0.0008+0.0016i] |
| $a_{33}$ =0.0144-0.0036i [-0.0053-0.0024i] | $a_{34}$ =-0.0264+0.0001i [-0.0103+0.0040i] |
| $a_{35}$ =-8.94*10$^{-9}$+3.04*10$^{-10}$i [2.17*10$^{-6}$-6.61*10$^{-7}$i] | $a_{36}$ =-9.83*10$^{-10}$+1.05*10$^{-11}$i [(-6.54+1.99i)*10$^{-6}$] |
| $a_{37}$ =-7.47*10$^{-11}$+8.59*10$^{-12}$i [(6.54-1.99i)*10$^{-6}$] | $a_{38}$ =-1.69*10$^{-9}$+7.38*10$^{-11}$i [-2.19*10$^{-6}$+6.63*10$^{-7}$i] |
| $a_{39}$ =6.85*10$^{-4}$-2.23*10$^{-5}$i [(-3.37+2.62i)*10$^{-4}$] | $a_{40}$ =(2.47+2.40i)*10$^{-4}$ [(-4.29+4.22i)*10$^{-4}$] |
| $a_{41}$ =-3.19*10$^{-5}$+1.33*10$^{-4}$i [8.76*10$^{-5}$-3.78*10$^{-6}$i] | $a_{42}$ =-2.18*10$^{-4}$-4.50*10$^{-5}$i [(4.24-6.04i)*10$^{-5}$] |
| $a_{43}$ =(-8.64-4.71i)*10$^{-5}$ [(1.38-1.34i)*10$^{-4}$] | $a_{44}$ =3.75*10$^{-6}$-2.47*10$^{-5}$i [(4.62-3.25i)*10$^{-5}$] |
| $a_{45}$ =-0.0027+0.0006i [(5.32-1.65i)*10$^{-4}$] | $a_{46}$ =-0.0014+0.0001i [5.51*10$^{-5}$+4.07*10$^{-4}$i] |
| $a_{47}$ =3.65*10$^{-4}$+6.14*10$^{-5}$i [-9.15*10$^{-5}$+1.23*10$^{-4}$i] | $a_{48}$ =-7.50*10$^{-6}$+4.94*10$^{-5}$i [(-9.12+9.36i)*10$^{-5}$] |
| $a_{49}$ =0.0056+0.0002i [0.0014-0.0006i] | $a_{50}$ =-4.65*10$^{-4}$-6.80*10$^{-5}$i [-1.32*10$^{-4}$-1.03*10$^{-5}$i] |
| $a_{51}$ =2.49*10$^{-7}$-7.61*10$^{-9}$i [1.41*10$^{-7}$-2.71*10$^{-8}$i] | $a_{52}$ =6.17*10$^{-8}$-2.37*10$^{-9}$i [1.83*10$^{-7}$-2.67*10$^{-8}$i] |

2. Mapping process

The previous step has shown how to obtain one knot or link in space. Fig. 5a and 5b show the trefoil knot. Here we provide the construction of trefoil knot in the lattice. We construct the lattice as

$$H_{def-tre} = \sum_{\langle i,j \rangle} t_{i,j} a_i^\dagger a_j + \sum_{\langle k,l \rangle} s_{k,l} a_k^\dagger a_l + \text{H.C.} \tag{A5}$$

The site in the lattice is often described by the integral coordinate. In Fig. 5c, the ranges of

$x$, $y$, $z$ coordinates in the lattice are $x \in [1,7]$, $y \in [1,8]$, $z \in [0,2]$, and the values of $x$, $y$, $z$ are all integers. Therefore, we seek all integral coordinates satisfying $|f_{deformed-tre}| \leq 5*10^{-9}$ and the sites described by these coordinates can comprise the deformed "discrete" trefoil knot. We list all these coordinates in sequence below,

**Table 2.** The coordinates $(x,y,z)$ satisfy $|f_{deformed-tre}| \leq 5*10^{-9}$.

| (2,3,2) | (2,4,1) | (3,5,1) | (4,6,1) | (5,6,2) | (6,5,2) | (7,4,2) | (6,3,2) | (5,3,1) |
|---|---|---|---|---|---|---|---|---|
| (4,4,1) | (4,5,2) | (3,6,2) | (4,7,2) | (5,8,2) | (6,7,2) | (6,6,1) | (5,5,1) | (5,4,2) |
| (4,3,2) | (3,2,2) | | | | | | | |

In the Hamiltonian, we set the coordinates shown in Table 2 to connect with the six adjacent sites through the coupling strength $t_{i,j} = 0.01$, and the coupling strengths between other sites are $s_{k,l} = 1$. We provide the construction of this Hamiltonian in the three-dimensional lattice in Fig. 5c. Red (blue) cylinders represent the coupling strength $t_{i,j}$ ($s_{k,l}$). We also show the distribution of localized eigenstates of $H$ in Fig. 5c. Red spheres label the sites occupied by the localized eigenstates. We find that the coordinates of these sites are exactly those shown in Table 2. To illustrate the distribution clearly, we use the purple tube in Fig. 5c to connect these sites and find that they comprise the trefoil knot. This trefoil knot is the "discrete" deformed trefoil knot in Fig. 5b.

3. Electric realization

Since the lattice contains three layers along the z direction, we use three printed circuit board (PCB) layers in the electric circuit. The detailed structure is provided in Fig. 5d. Every two nearest neighboring nodes in the circuit is connected through the capacitor $C_1$ ($C_2$), which corresponds to the coupling strength $t_{i,j}$ ($s_{k,l}$) in the lattice. In the electric design, the dynamics of this circuit can be described by one Laplacian $J$ which bridges the current and voltages in the circuit. By choosing appropriate grounding inductors, we can make the Laplacian $J$ similar to the Hamiltonian $H_{def-tre}$. For three nodes in the circuit (contained in

the orange dotted triangle) in Fig. 5d, the Laplacian is addressed as

$$J = \begin{pmatrix} \cdots & \cdots & & \cdots \\ 6i\omega C_2 + 6(i\omega L_2)^{-1} & i\omega C_2 & 0 & \\ \cdots & i\omega C_2 & 5i\omega C_2 + 5(i\omega L_2)^{-1} + i\omega C_1 + (i\omega L_1)^{-1} & i\omega C_1 & \cdots \\ & 0 & i\omega C_1 & 6i\omega C_1 + 6(i\omega L_1)^{-1} \\ \cdots & \cdots & & \cdots \end{pmatrix} \quad (A6)$$

In our design, when we set $\omega = \frac{1}{\sqrt{L_1 C_1}} = \frac{1}{\sqrt{L_2 C_2}}$, the diagonal elements in the matrix $J$ are eliminated, and the Laplacian changes to the expression shown in the inset of Fig. 5d.

In our experiment, we use the impedance as the measured quantity. The impedance between $a$th node and $b$th node is $Z_{a,b} = \frac{V_a - V_b}{I_{a,b}} = \sum_{j_n \neq 0} \frac{|\varphi_{n,a} - \varphi_{n,b}|^2}{j_n}$. Here, $V_a$ ($V_b$) is the electric potential at $a(b)$th node and $I_{a,b}$ is the current between $a$th node and $b$th node. The symbols $j_n$ and $\varphi_n$ represent the $n$th admittance eigenvalue and the corresponding eigenstate of $J$. For convenience, we set the $b$th node as the ground. When the node is connected by the capacitor $C_1$ ($C_2$), the corresponding term in Laplacian $J$ is $i\omega C_1$ ($i\omega C_2$). When the electric circuit shown in Fig. 5d is designed, we can obtain the distribution of impedance which is similar to that of localized eigenstates in Fig. 5c. The experimental distributions of impedance have been provided in the main text.

After completing these three steps, we realize the "discrete" trefoil knot in the electric circuit. To make this realization clear, we connect these "special" nodes having large impedances together. Since these nodes distribute at the diagonal nodes of each plaquette in the circuit (red spheres in Fig. 1a and large spheres in Fig. 1f of the main text), we connect these diagonal nodes of each plaquette in the circuit in sequence. The obtained connection is exactly the trefoil knot (see Fig. 1a and Fig. 1f in the main text). Compared with other experimental realizations of knots, the realization of trefoil knot in the electric circuit is easy and controllable. Moreover, not only the trefoil knot is implemented electrically, other knots and links can also be realized in the electric circuit in the same way.

**Appendix B: The experimental details in the electric circuit**

Here, we describe the experimental details in the electric circuit. In our experiment, we choose the electric capacitors CC41-0603-CG-50V-100pF-F(N) and CC41-0603-CG-50V-10nF-F(N), which are described as the small capacitors in red ($C_1$) and the large capacitors in blue ($C_2$) in the main text, respectively. The types of inductors are NLV32T-3R3J-EF and NLV32T-033J-EF. We measure the impedance at each node by connecting the impedance analyzer to the measurement connectors. The cooper pillars are connected to the grounding inductors, which also sustain the three PCB layers.

We use the WK6500B impedance analyzer to measure the impedance. In principle, we need to measure the impedance between the nodes and the ground at the resonant frequency $\omega_0 = 1/\sqrt{L_1 C_1}$. Due to the existence of various errors in the experiments, the measured frequency is a little smaller than the resonant frequency. To obtain the appropriate results, we use the impedance analyzer to sweep the frequencies around the resonant frequency. We choose the certain frequency where the peak of impedance appears and we use this peak value as the impedance of this node.

In our experiment, the values of electric inductors and capacitors, of course, are not ideal, and there exists spurious inductive coupling in the experimental setup, but the dominant error is from the connecting wires. The type of these wires is DB9. These wires are used to realize cyclic boundary condition and connect different PCB layers. The parasitic inductance from long connecting wires cannot be neglected. When we measure the impedance of nodes connecting with adjacent nodes through the small electric capacitor $C_1$, the impedance of these nodes are very large in the ideal circuit simulation. However, in fact, due to the non-negligible parasitic inductance of the long wires in the experiment, we find that the impedances for these nodes at the edge of the PCB layer are often a little bit smaller than those at the inner part of the layer, but are still much larger than the nodes with small impedances. So this error does not ruin our experimental results.

**Appendix C: The realization of change process in the lattice and electric circuit**

Here, we describe how to realize the change process in the electric circuit. Similar to the construction in Appendix A, we firstly provide the functions describing the curves from Figs.

2a to 2e of the main text, then we give the Hamiltonians of lattices corresponding to the electric designs from Figs. 2f to 2j of the main text.

The functions from Figs.2a to 2e in the main text are listed below in sequence:

$$\begin{cases} f_{a,purple} = (x-1)^2 + (y-2)^2 - 1 = 0; x \in [1,4], y \in [1,3], z = 2 \\ f_{a,orange} = (x-4)^2 + (z-3)^2 - 1 = 0; x \in [1,4], y = 2, z \in [1,4] \end{cases}, \quad (C1)$$

$$\begin{cases} f_{b,purple} = (x-1)^2 + (y-2)^2 - 1 = 0; x \in [1,4], y \in [1,3], z = 2 \\ f_{b,orange} = (x-4)^2 + (z-2)^2 - 1 = 0; x \in [1,4], y = 2, z \in [1,4] \end{cases}, \quad (C2)$$

$$\begin{cases} f_{c,purple} = (x-1)^2 + (y-2)^2 - 1 = 0; x \in [1,4], y \in [1,3], z = 2 \\ f_{c,orange} = (x-3)^2 + (z-2)^2 - 1 = 0; x \in [1,3], y = 2, z \in [1,4] \end{cases}, \quad (C3)$$

$$\begin{cases} f_{d,purple} = (x-2)^2 + (y-2)^2 - 1 = 0; x \in [2,4], y \in [1,3], z = 2 \\ f_{d,orange} = (x-3)^2 + (z-2)^2 - 1 = 0; x \in [1,3], y = 2, z \in [1,4] \end{cases}, \quad (C4)$$

$$\begin{cases} f_{e,purple} = (x-3)^2 + (y-2)^2 - 1 = 0; x \in [3,4], y \in [1,3], z = 2 \\ f_{e,orange} = (x-3)^2 + (z-2)^2 - 1 = 0; x \in [1,3], y = 2, z \in [1,4] \end{cases}. \quad (C5)$$

Here we provide the constructions of these five structures in the lattices. The ranges of $x$, $y$, $z$ coordinates in the lattices are $x \in [1,4]$, $y \in [1,3]$, $z \in [1,4]$, and the values of $x$, $y$, $z$ are all integers. We construct the lattices as

$$H_{i=a,b,c,d,e} = \sum_{\langle i,j \rangle} t_{i,j} a_i^\dagger a_j + \sum_{\langle k,l \rangle} s_{k,l} a_k^\dagger a_l + \text{H.C.} \quad (C6)$$

The sites in the lattices are often described by the integral coordinates. Therefore, we seek all integral coordinates satisfying $|f_{i,purple}| = 0$ and $|f_{i,orange}| = 0$. We list all these coordinates in sequence below,

**Table 3.** The coordinates $(x, y, z)$ satisfy $|f_{i,purple}| = 0$ and $|f_{i,orange}| = 0$

|       | $|f_{i,purple}| = 0$          | $|f_{i,orange}| = 0$          |
|-------|-------------------------------|-------------------------------|
| $i=a$ | (1,1,2), (2,2,2), (1,3,2)     | (4,2,2), (3,2,3), (4,2,4)     |
| $i=b$ | (1,1,2), (2,2,2), (1,3,2)     | (4,2,1), (3,2,2), (4,2,3)     |
| $i=c$ | (1,1,2), (2,2,2), (1,3,2)     | (3,2,1), (2,2,2), (3,2,3)     |
| $i=d$ | (2,1,2), (3,2,2), (2,3,2)     | (3,2,1), (2,2,2), (3,2,3)     |
| $i=e$ | (3,1,2), (4,2,2), (3,3,2)     | (3,2,1), (2,2,2), (3,2,3)     |

In the Hamiltonian, we set the coordinates shown in Table 3 to connect with the six adjacent sites through the coupling strength $t_{i,j} = 0.01$, and the coupling strengths between other sites are $s_{k,l} = 1$. We provide the construction of these Hamiltonians in these three-dimensional lattices from Figs. 6a to Fig. 6e. Red (blue) cylinders represent the coupling strength $t_{i,j}$ ($s_{k,l}$). We also show the distribution of localized eigenstates of $H$ from Fig. 6a to Fig. 6e.

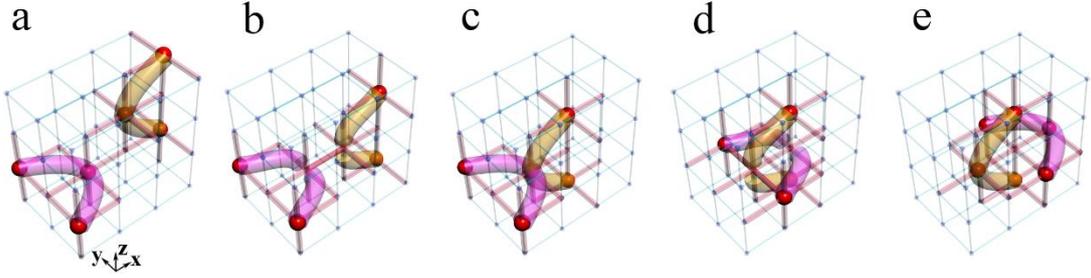

**Fig. 6. Constructions of the geometric structures shown from Fig. 2a to Fig. 2e in the main text.** Red (blue) cylinders represent the couplings strengths $t = 0.01$ ($s = 1$). Red spheres in the lattices are the sites occupied by the localized eigenstates.

Red spheres label the sites occupied by the localized eigenstates. We find that for the cases $i = a, c, e$, the coordinates of these sites are exactly those shown in Table 3. But for the case $i = b$, the sites with coordinates (2,2,2) in $|f_{b,purple}| = 0$ and (3,2,2) in $|f_{b,orange}| = 0$ are not occupied by the localized eigenstates; and for the case $i = d$, the sites with coordinates (3,2,2) in $|f_{b,purple}| = 0$ and (2,2,2) in $|f_{b,orange}| = 0$ are not occupied by the localized eigenstates. So if we connect the sites occupied by the localized eigenstates, we can obtain the corresponding geometric structures depicted in Figs. 2a, 2c and 2e in the main text, but not obtain the corresponding geometric structures depicted in Figs. 2b and 2d in the main text. This is because, in the lattice, when the sites connect with neighboring sites by $t_{i,j}$ or $s_{k,l}$, the hopping rate between sites shown in Hamiltonian is $t_{i,j}$ or $s_{k,l}$. Consider $t_{i,j} < s_{k,l}$, so electrons are more likely trapped at the site connected to neighboring sites by $t_{i,j}$. But when two sites connecting with neighboring sites by $t_{i,j}$ connect directly, electrons are not limited to be at one of these two sites, which does not bring strong localization at these two sites simultaneously. If we still connect the sites occupied by localized eigenstates, we can find the connections are broken apart in the middle, see Figs. 6b and 6d. So we cannot recover the geometric structures in Figs. 2b and 2d of the main text, respectively. It means that during the change of one crossing between two tubes, the sites occupied by localized eigenstates cannot correspond to the geometric structures twice. It is noted that this phenomenon does not have any influence on counting the times of changing crossings in the lattice.

Consider the electric realizations correspond to the lattices, the distributions of impedance can exactly recover the geometric structures for cases $i=a,c,e$, but not form the structures for cases $i=b,d$. The electric simulation results of impedances for cases $i=a,b,c,d,e$ are provided from Fig. 2f to Fig. 2j in the main text.

**Appendix D: The observation of unknotting numbers for trefoil knot, figure-8 knot and $8_3$ knot in the electric circuits.**

**a. Trefoil knot**

When we change one crossing in the trefoil knot, the trefoil knot can change to the unknot. Here we provide the realizations of structures during the change process from Fig. 7a to Fig. 7e.

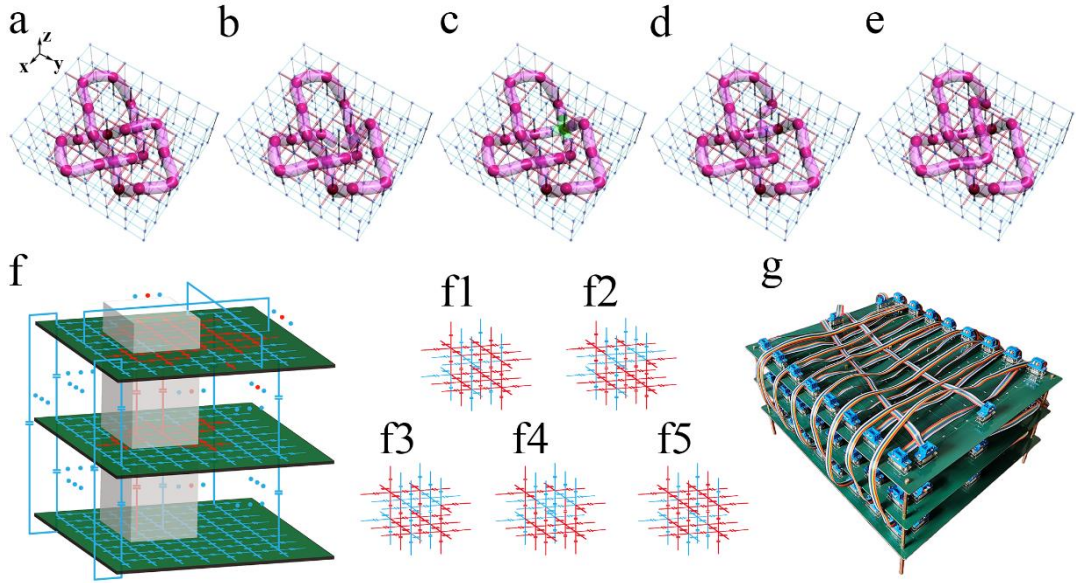

**Fig. 7. Constructions to show the continuous change process from the trefoil knot to unknot.** From (a) to (e), we illustrate the constructions from the trefoil knot to unknot in the lattices. Red (blue) cylinders represent the coupling strengths $t=0.01$ ($s=1$). Red spheres in the lattices are the sites occupied by the localized eigenstates. (f) Electric designs to realize the five lattices from (a) to (e). The gray regions are different for five structures from (a) to (e). Details in the gray region for (a) to (e) are provided in (f1) to (f5), respectively. The value of capacitors in red (blue) is $C_1$=100pF ($C_2$=10nF). (g) The experimental setup to realize these five structures.

Constructions of these structures are similar to the description in Appendix A. These five structures from Fig. 7a to Fig. 7e are constructed in the lattices as Eq. (D1),

$$H_{un} = \sum_{\langle i,j \rangle} t_{i,j} a_i^\dagger a_j + \sum_{\langle k,l \rangle} s_{k,l} a_k^\dagger a_l + \text{H.C.} \qquad (D1)$$

The first structure is to form the trefoil knot and the fifth structure is to form the unknot. The construction for the first structure has been provided in Appendix A. The functions for the four structures from Fig. 7b to Fig. 7e can be obtained in a similar way as the deformed trefoil knot (Eq. (A4)). Consider the lengthy expressions of these functions for these four structures, we do not provide the details of functions here. Since the coordinates in the lattice are integers, we provide the coordinates satisfying the corresponding functions in Table 4. In our design, the coordinates in Table 4 connect with the six adjacent sites through the coupling strength $t_{i,j} = 0.01$, and the coupling strengths between other sites are $s_{k,l} = 1$.

**Table 4.** The coordinates $(x, y, z)$ connect with all adjacent sites through $t_{i,j} = 0.01$

| The second structure in Fig. 7b | | | | | | | | |
|---|---|---|---|---|---|---|---|---|
| (2,3,2) | (2,4,1) | (3,5,1) | (4,6,1) | (5,6,2) | (6,5,2) | (7,4,2) | (6,3,2) | (5,3,1) |
| (4,4,1) | (4,5,1) | (3,6,2) | (4,7,2) | (5,8,2) | (6,7,2) | (6,6,1) | (5,5,1) | (5,4,2) |
| (4,3,2) | (3,2,2) | | | | | | | |
| The third structure in Fig. 7c | | | | | | | | |
| (2,3,2) | (2,4,1) | (3,5,1) | (4,6,1) | (5,6,2) | (6,5,2) | (7,4,2) | (6,3,2) | (5,3,1) |
| (4,4,1) | (3,5,1) | (3,6,2) | (4,7,2) | (5,8,2) | (6,7,2) | (6,6,1) | (5,5,1) | (5,4,2) |
| (4,3,2) | (3,2,2) | | | | | | | |
| The fourth structure in Fig. 7d | | | | | | | | |
| (2,3,2) | (2,4,1) | (3,4,1) | (4,5,1) | (5,6,2) | (6,5,2) | (7,4,2) | (6,3,2) | (5,3,1) |
| (4,4,1) | (3,5,1) | (3,6,2) | (4,7,2) | (5,8,2) | (6,7,2) | (6,6,1) | (5,5,1) | (5,4,2) |

| (4,3,2) | (3,2,2) | | | | | | | |
|---------|---------|---|---|---|---|---|---|---|
| The fifth structure (the unknot in Fig. 1b of the main text) in Fig. 7e | | | | | | | | |
| (2,3,2) | (2,4,1) | (3,4,2) | (4,5,2) | (5,6,2) | (6,5,2) | (7,4,2) | (6,3,2) | (5,3,1) |
| (4,4,1) | (3,5,1) | (3,6,2) | (4,7,2) | (5,8,2) | (6,7,2) | (6,6,1) | (5,5,1) | (5,4,2) |
| (4,3,2) | (3,2,2) | | | | | | | |

Comparing the chosen coordinates in Table 4 to realize the five structures in the lattices, we find that the coordinates from the first to the fifth structure are changed as: first (Fig. 7a)→second (Fig. 7b), (4,5,2)→(4,5,1); second (Fig. 7b)→third (Fig. 7c), (4,5,1)→(3,5,1); third (Fig. 7c)→fourth (Fig. 7d), (3,5,1)→(3,4,1), (4,6,1)→(4,5,1); and fourth (Fig. 7d)→fifth (Fig. 7e), (3,4,1)→(3,4,2), (4,5,1)→(4,5,2). In the change process, only one or two sites coupled to neighboring sites by strength $t_{i,j} = 0.01$ are changed and these sites are nearest neighboring to each other at any two adjacent steps. In this sense, we can view this change process continuously. It means that we need only continuously to modulate some coupling strengths at each step in the change process. Red spheres from Fig. 7a to Fig. 7e are the occupations of localized eigenstates. We can find that for the first trefoil knot (Fig. 7a), the third structure (Fig. 7c) and the fifth unknot (Fig. 7e), the coordinates of sites occupied by the localized eigenstates are exactly those presented in Table 2 and 4; but for the second (Fig. 7b) and fourth (Fig. 7d) structures, some sites with coordinates shown in Table 4 are not occupied by localized eigenstates. This phenomenon corresponds to the description of changing one crossing in the text around Fig. 6. Moreover, if we change the structures continuously following the sequence as, first (Fig. 7a), second (Fig. 7b), third (Fig. 7c), second (Fig. 7b) and first (Fig. 7a) structures, there are also two cases where the connections formed by localized eigenstates cannot recover the geometric structures. During this change, the trefoil return back to itself finally. Fortunately, the definition of unknotting number is the least time of crossing change necessary to change the knot into an unknot [1]. So the change making trefoil return back to itself is meaningless for the invariant unknotting number.

The corresponding electric realizations are presented in Fig. 7f. Similar to the electric realization in Appendix A, we connect every two nearest neighboring nodes in the electric circuit through the capacitors. The nodes with the coordinates presented in Table 4 are connected by the small electric capacitor $C_1$=100pF. Other nodes are connected by the large capacitor $C_2$=10nF. Consider the correspondence between the lattice and the electric circuit, we need to change some capacitors and their associated grounding inductors at each step. The corresponding electric setup is presented in Fig. 7g.

In Fig. 8, we provide the measurement outcomes of distributions of impedances for the corresponding electric designs in Fig. 7 (f1 to f5).

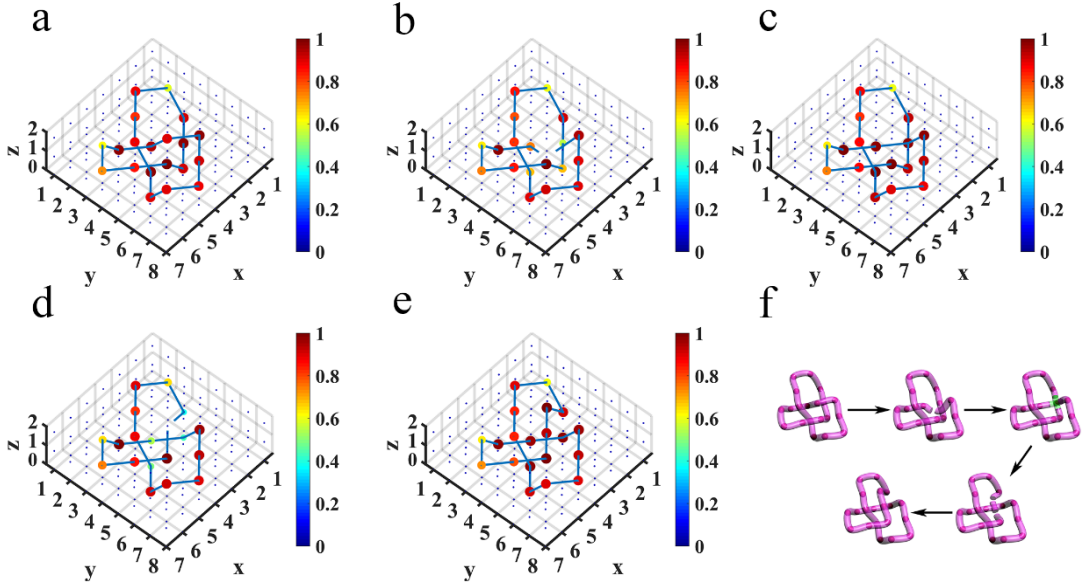

**Fig. 8. Experimental distributions of impedance for the five structures from trefoil knot to unknot.** From (a) to (e), the value of impedance at each node has been normalized to the maximum value. (f) The structures recovered from the distributions of large impedances during the change in the circuit.

From the distributions of impedance in Fig. 8, we can find that the nodes connected by small electric capacitors $C_1$ are exactly having a large impedance for the first, third and fifth cases (Figs. 8a, 8c and 8e). But for the second and fourth cases (Figs. 8b and 8d), some nodes connected by the capacitors $C_1$ are not possessing large impedances. So these experimental results are consistent with those in the simulations.

**b. Figure-8 knot**

Not only the change of crossing once in the trefoil knot can be revealed in the electric circuit, but changes of crossings in other knots can also be provided. The construction details in the figure-8 knot and $8_3$ knot are the same as the electrical trefoil knot above. We firstly provide the constructions of figure-8 knot and $8_3$ knot in lattices, then map to the electric circuits. The Hamiltonian to realize the change from figure-8 knot to unknot in the lattice is shown as,

$$H_{figure-8} = \sum_{\langle i,j \rangle} t_{i,j} a_i^\dagger a_j + \sum_{\langle k,l \rangle} s_{k,l} a_k^\dagger a_l + \text{H.C.} \qquad (D2)$$

Here, we do not provide the details of functions for these structures from figure-8 knot to unknot. Consider the coordinates in the lattices are integers, we only provide the integral coordinates satisfying the corresponding functions in Table 5. To realize the construction from the figure-8 knot to unknot, we show the coordinates in Table 5 that connect with the six adjacent sites through the coupling strength $t_{i,j} = 0.01$, and the coupling strengths between other sites are $s_{k,l} = 1$.

**Table 5.** The coordinates $(x, y, z)$ connect with all adjacent sites through $t_{i,j} = 0.01$

| The first structure (figure-8 knot) in Fig. 9a | | | | | | | | | |
|---|---|---|---|---|---|---|---|---|---|
| (2,2,2) | (2,3,3) | (2,4,2) | (2,5,3) | (3,6,3) | (4,7,3) | (5,6,3) | (6,6,2) | (5,6,1) | (4,5,1) | (4,4,2) |
| (3,3,2) | (3,2,3) | (3,1,2) | (4,1,1) | (5,2,1) | (5,3,2) | (6,4,2) | (5,5,2) | (4,6,2) | (5,7,2) | (6,7,1) |
| (7,7,2) | (7,6,1) | (7,5,2) | (7,4,1) | (7,3,2) | (6,2,2) | (5,1,2) | (4,2,2) | (3,2,1) | | |
| The second structure in Fig. 9b | | | | | | | | | |
| (2,2,2) | (2,3,3) | (2,4,2) | (2,5,3) | (3,6,3) | (4,7,3) | (5,6,3) | (6,6,2) | (5,6,1) | (4,5,1) | (4,4,2) |
| (3,3,2) | (3,2,3) | (3,1,2) | (4,1,1) | (5,2,1) | (5,3,2) | (6,4,2) | (5,5,2) | (4,6,2) | (5,7,2) | (6,7,1) |
| (7,7,2) | (7,6,1) | (7,5,2) | (7,4,1) | (7,3,2) | (6,2,2) | (5,1,2) | (4,2,2) | (3,2,2) | | |
| The third structure in Fig. 9c | | | | | | | | | |
| (2,2,2) | (2,3,3) | (2,4,2) | (2,5,3) | (3,6,3) | (4,7,3) | (5,6,3) | (6,6,2) | (5,6,1) | (4,5,1) | (4,4,2) |
| (3,3,2) | (3,2,3) | (3,1,2) | (4,1,1) | (5,2,1) | (5,3,2) | (6,4,2) | (5,5,2) | (4,6,2) | (5,7,2) | (6,7,1) |
| (7,7,2) | (7,6,1) | (7,5,2) | (7,4,1) | (7,3,2) | (6,2,2) | (5,1,2) | (4,2,2) | (3,1,2) | | |
| The fourth structure in Fig. 9d | | | | | | | | | |
| (2,2,2) | (2,3,3) | (2,4,2) | (2,5,3) | (3,6,3) | (4,7,3) | (5,6,3) | (6,6,2) | (5,6,1) | (4,5,1) | (4,4,2) |

| (3,3,2) | (3,2,2) | (3,1,2) | (4,1,1) | (5,2,1) | (5,3,2) | (6,4,2) | (5,5,2) | (4,6,2) | (5,7,2) | (6,7,1) |
| --- | --- | --- | --- | --- | --- | --- | --- | --- | --- | --- |
| (7,7,2) | (7,6,1) | (7,5,2) | (7,4,1) | (7,3,2) | (6,2,2) | (5,1,2) | (4,2,2) | (3,1,2) | | |
| The fifth structure in Fig. 9e | | | | | | | | | | |
| (2,2,2) | (2,3,3) | (2,4,2) | (2,5,3) | (3,6,3) | (4,7,3) | (5,6,3) | (6,6,2) | (5,6,1) | (4,5,1) | (4,4,2) |
| (3,3,2) | (3,2,1) | (3,2,1) | (4,1,1) | (5,2,1) | (5,3,2) | (6,4,2) | (5,5,2) | (4,6,2) | (5,7,2) | (6,7,1) |
| (7,7,2) | (7,6,1) | (7,5,2) | (7,4,1) | (7,3,2) | (6,2,2) | (5,1,2) | (4,2,2) | (3,1,2) | | |

When compared with the chosen coordinates to realize the five structures in the lattices, in Table 5, we find that the coordinates from the first to the fifth structure are changed as: first (Fig. 9a)→second (Fig. 9b), (3,2,1)→(3,2,2); second (Fig. 9b)→third (Fig. 9c), (3,2,2)→(3,1,2); third (Fig. 9c)→fourth (Fig. 9d), (3,2,3)→(3,2,2); and fourth (Fig. 9d)→fifth (Fig. 9e), (3,2,2)→(3,2,1), (3,1,2)→(3,2,1). In the change process, only one or two sites coupled to neighboring sites by strength $t_{i,j} = 0.01$ are changed and these sites are nearest neighboring to each other at any two adjacent steps. In this sense, we can view this change process continuously. It means that we need only continuously to modulate some coupling strengths at each step in the change process. The corresponding realizations in lattices are presented in Fig. 9.

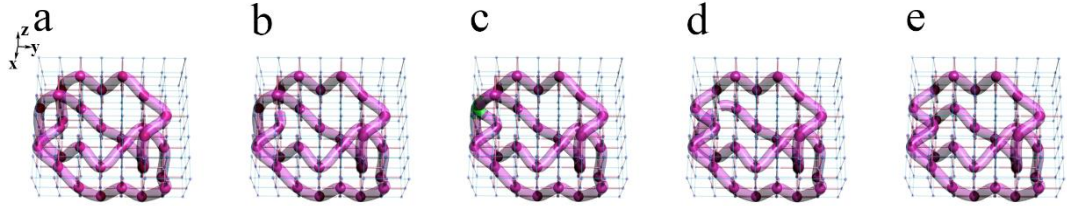

**Fig. 9. Constructions from figure-8 knot to unknot in the lattices.** From (a) to (e), the figure-8 knot changes to unknot. Red (blue) cylinders represent the coupling strengths $t = 0.01$ ($s = 1$). Red spheres in the lattices are the sites occupied by the localized eigenstates. These red spheres are connected in purple tubes.

Red spheres in Fig. 9 represent the sites occupied by the localized eigenstates. We can find that for the first figure-8 knot (Fig. 9a), the third structure (Fig. 9c) and the fifth unknot (Fig. 9e), the coordinates of sites occupied by the localized eigenstates are exactly those presented in Table 5; but for the second (Fig. 9b) and fourth (Fig. 9d) structures, some sites with coordinates shown in Table 5 are not occupied by localized eigenstates. This phenomenon corresponds to the description of changing one crossing in the text around Fig. 6.

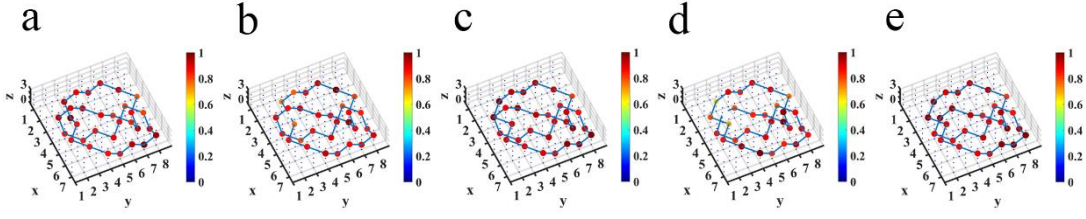

**Fig. 10. Simulated distributions of impedance for the five structures from figure-8 knot to unknot.** From (a) to (e), the value of impedance at each node has been normalized to the maximum value.

Similar to the electric realization above, we connect every two nearest neighboring nodes in the electric circuit through the capacitors, and the nodes with the coordinates presented in Table 5 are connected by small capacitors $C_1=100\text{pF}$. Other nodes are connected by large capacitors $C_2=10\text{nF}$. Consider the correspondence between the lattice and the electric circuit, we only need to change some capacitors and their associated grounding inductors at each step. In Fig. 10, we provide the distributions of impedances. We can find that for the first figure-8 knot (Fig. 10a), the third structure (Fig. 10c) and the fifth unknot (Fig. 10e), the coordinates of nodes possessing large impedance are exactly those presented in Table 5; but for the second (Fig. 10b) and fourth (Fig. 10d) structures, some sites with coordinates shown in Table 5 do not have large impedances, and the connections are broken apart. This corresponds to the observation of localization of eigenstates in the lattices in Fig. 9.

### c. $8_3$ knot

To show the change process that changes from $8_3$ knot to unknot, we need to change crossing in $8_3$ knot twice. The Hamiltonian to realize the change from the $8_3$ knot to unknot in lattice is shown as

$$H_{8_3 knot} = \sum_{\langle i,j \rangle} t_{i,j} a_i^\dagger a_j + \sum_{\langle k,l \rangle} s_{k,l} a_k^\dagger a_l + \text{H.C.} \qquad (D3)$$

Here, we do not provide the details of functions for these structures from the $8_3$ knot to unknot. Consider the coordinates in the lattices are integers, we only provide the integral coordinates satisfying the corresponding functions in Table 6 and 7. To realize the construction of $8_3$ knot to unknot, we show the coordinates in Table 6 that connect with the six adjacent sites through the coupling strength $t_{i,j} = 0.01$ for the first five structures in Fig. 11, and the coupling strengths between other sites are $s_{k,l} = 1$. When obtaining the fifth structure, we have changed the crossing in the $8_3$ knot once. At this time, the obtained fifth structure is still knotted.

**Table 6.** The coordinates $(x, y, z)$ connect with all adjacent sites through $t_{i,j} = 0.01$

| The first structure ($8_3$ knot) in Fig. 11a | | | | | | | | | |
|---|---|---|---|---|---|---|---|---|---|
| (2,4,2) | (3,5,2) | (2,6,2) | (3,7,2) | (2,8,2) | (3,9,2) | (2,10,2) | (3,11,2) | (2,12,2) | (3,13,2) | (2,14,2) |
| (2,15,3) | (2,16,2) | (3,17,2) | (4,18,2) | (4,19,1) | (5,20,1) | (6,20,2) | (7,20,3) | (8,19,3) | (8,18,2) | (9,18,1) |
| (10,17,1) | (10,16,2) | (10,15,3) | (10,14,2) | (9,13,2) | (10,12,2) | (9,11,2) | (10,10,2) | (9,9,2) | (10,8,2) | (9,7,2) |
| (10,6,2) | (9,5,2) | (10,4,2) | (9,3,2) | (8,2,2) | (7,2,3) | (6,3,3) | (5,4,3) | (5,5,2) | (5,6,1) | (6,7,1) |
| (7,8,1) | (7,9,2) | (7,10,3) | (6,11,3) | (5,12,3) | (5,13,2) | (5,14,1) | (6,15,1) | (7,16,1) | (8,16,2) | (8,17,3) |
| (9,18,3) | (10,19,3) | (10,20,2) | (9,21,2) | (8,20,2) | (7,19,2) | (6,18,2) | (5,19,2) | (4,20,2) | (3,21,2) | (2,20,2) |
| (2,19,1) | (3,18,1) | (4,17,1) | (4,16,2) | (5,16,3) | (6,15,3) | (7,14,3) | (7,13,2) | (7,12,1) | (6,11,1) | (5,10,1) |
| (5,9,2) | (5,8,3) | (6,7,3) | (7,6,3) | (7,5,2) | (7,4,1) | (6,3,1) | (5,2,1) | (4,2,2) | (3,3,2) | |
| The second structure in Fig. 11b | | | | | | | | | |
| (2,4,2) | (3,5,2) | (2,6,2) | (3,7,2) | (2,8,2) | (3,9,2) | (2,10,2) | (3,11,2) | (2,12,2) | (3,13,2) | (2,14,2) |
| (2,15,3) | (2,16,2) | (3,17,2) | (4,18,2) | (4,19,1) | (5,20,1) | (6,20,2) | (7,20,3) | (8,19,3) | (8,18,2) | (9,18,1) |
| (10,17,1) | (10,16,2) | (10,15,3) | (10,14,2) | (9,13,2) | (10,12,2) | (9,11,2) | (10,10,2) | (9,9,2) | (10,8,2) | (9,7,2) |
| (10,6,2) | (9,5,2) | (10,4,2) | (9,3,2) | (8,2,2) | (7,2,3) | (6,3,3) | (5,4,3) | (5,5,2) | (5,6,1) | (6,7,1) |
| (7,8,1) | (7,9,2) | (7,10,3) | (6,11,3) | (5,12,3) | (5,13,2) | (5,14,2) | (6,15,2) | (7,16,2) | (8,16,2) | (8,17,3) |
| (9,18,3) | (10,19,3) | (10,20,2) | (9,21,2) | (8,20,2) | (7,19,2) | (6,18,2) | (5,19,2) | (4,20,2) | (3,21,2) | (2,20,2) |
| (2,19,1) | (3,18,1) | (4,17,1) | (4,16,2) | (5,16,3) | (6,15,3) | (7,14,3) | (7,13,2) | (7,12,1) | (6,11,1) | (5,10,1) |
| (5,9,2) | (5,8,3) | (6,7,3) | (7,6,3) | (7,5,2) | (7,4,1) | (6,3,1) | (5,2,1) | (4,2,2) | (3,3,2) | |
| The third structure in Fig. 11c | | | | | | | | | |
| (2,4,2) | (3,5,2) | (2,6,2) | (3,7,2) | (2,8,2) | (3,9,2) | (2,10,2) | (3,11,2) | (2,12,2) | (3,13,2) | (2,14,2) |
| (2,15,3) | (2,16,2) | (3,17,2) | (4,18,2) | (4,19,1) | (5,20,1) | (6,20,2) | (7,20,3) | (8,19,3) | (8,18,2) | (9,18,1) |
| (10,17,1) | (10,16,2) | (10,15,3) | (10,14,2) | (9,13,2) | (10,12,2) | (9,11,2) | (10,10,2) | (9,9,2) | (10,8,2) | (9,7,2) |

| | | | | | | | | | |
|---|---|---|---|---|---|---|---|---|---|
| (10,6,2) | (9,5,2) | (10,4,2) | (9,3,2) | (8,2,2) | (7,2,3) | (6,3,3) | (5,4,3) | (5,5,2) | (5,6,1) | (6,7,1) |
| (7,8,1) | (7,9,2) | (7,10,3) | (6,11,3) | (5,12,3) | (5,13,2) | (5,14,3) | (6,15,3) | (7,16,3) | (8,16,2) | (8,17,3) |
| (9,18,3) | (10,19,3) | (10,20,2) | (9,21,2) | (8,20,2) | (7,19,2) | (6,18,2) | (5,19,2) | (4,20,2) | (3,21,2) | (2,20,2) |
| (2,19,1) | (3,18,1) | (4,17,1) | (4,16,2) | (5,16,3) | (6,15,3) | (7,14,3) | (7,13,2) | (7,12,1) | (6,11,1) | (5,10,1) |
| (5,9,2) | (5,8,3) | (6,7,3) | (7,6,3) | (7,5,2) | (7,4,1) | (6,3,1) | (5,2,1) | (4,2,2) | (3,3,2) | |

| The fourth structure in Fig. 11d | | | | | | | | | |
|---|---|---|---|---|---|---|---|---|---|
| (2,4,2) | (3,5,2) | (2,6,2) | (3,7,2) | (2,8,2) | (3,9,2) | (2,10,2) | (3,11,2) | (2,12,2) | (3,13,2) | (2,14,2) |
| (2,15,3) | (2,16,2) | (3,17,2) | (4,18,2) | (4,19,1) | (5,20,1) | (6,20,2) | (7,20,3) | (8,19,3) | (8,18,2) | (9,18,1) |
| (10,17,1) | (10,16,2) | (10,15,3) | (10,14,2) | (9,13,2) | (10,12,2) | (9,11,2) | (10,10,2) | (9,9,2) | (10,8,2) | (9,7,2) |
| (10,6,2) | (9,5,2) | (10,4,2) | (9,3,2) | (8,2,2) | (7,2,3) | (6,3,3) | (5,4,3) | (5,5,2) | (5,6,1) | (6,7,1) |
| (7,8,1) | (7,9,2) | (7,10,3) | (6,11,3) | (5,12,3) | (5,13,2) | (5,14,3) | (6,15,3) | (7,16,3) | (8,16,2) | (8,17,3) |
| (9,18,3) | (10,19,3) | (10,20,2) | (9,21,2) | (8,20,2) | (7,19,2) | (6,18,2) | (5,19,2) | (4,20,2) | (3,21,2) | (2,20,2) |
| (2,19,1) | (3,18,1) | (4,17,1) | (4,16,2) | (5,16,2) | (6,15,2) | (7,14,2) | (7,13,2) | (7,12,1) | (6,11,1) | (5,10,1) |
| (5,9,2) | (5,8,3) | (6,7,3) | (7,6,3) | (7,5,2) | (7,4,1) | (6,3,1) | (5,2,1) | (4,2,2) | (3,3,2) | |

| The fifth structure in Fig. 11e | | | | | | | | | |
|---|---|---|---|---|---|---|---|---|---|
| (2,4,2) | (3,5,2) | (2,6,2) | (3,7,2) | (2,8,2) | (3,9,2) | (2,10,2) | (3,11,2) | (2,12,2) | (3,13,2) | (2,14,2) |
| (2,15,3) | (2,16,2) | (3,17,2) | (4,18,2) | (4,19,1) | (5,20,1) | (6,20,2) | (7,20,3) | (8,19,3) | (8,18,2) | (9,18,1) |
| (10,17,1) | (10,16,2) | (10,15,3) | (10,14,2) | (9,13,2) | (10,12,2) | (9,11,2) | (10,10,2) | (9,9,2) | (10,8,2) | (9,7,2) |
| (10,6,2) | (9,5,2) | (10,4,2) | (9,3,2) | (8,2,2) | (7,2,3) | (6,3,3) | (5,4,3) | (5,5,2) | (5,6,1) | (6,7,1) |
| (7,8,1) | (7,9,2) | (7,10,3) | (6,11,3) | (5,12,3) | (5,13,2) | (5,14,3) | (6,15,3) | (7,16,3) | (8,16,2) | (8,17,3) |
| (9,18,3) | (10,19,3) | (10,20,2) | (9,21,2) | (8,20,2) | (7,19,2) | (6,18,2) | (5,19,2) | (4,20,2) | (3,21,2) | (2,20,2) |
| (2,19,1) | (3,18,1) | (4,17,1) | (4,16,2) | (5,16,1) | (6,15,1) | (7,14,1) | (7,13,2) | (7,12,1) | (6,11,1) | (5,10,1) |
| (5,9,2) | (5,8,3) | (6,7,3) | (7,6,3) | (7,5,2) | (7,4,1) | (6,3,1) | (5,2,1) | (4,2,2) | (3,3,2) | |

When compared with the chosen coordinates to realize the five structures in the lattices, in Table 6, we find that the coordinates from the first to the fifth structure are changed as: first (Fig. 11a)→second (Fig. 11b), (5,14,1)→(5,14,2), (6,15,1)→(6,15,2), (7,16,1)→(7,16,2); second (Fig. 11b)→third (Fig. 11c), (5,14,2)→(5,14,3), (6,15,2)→(6,15,3), (7,16,2)→(7,16,3); third (Fig. 11c)→fourth (Fig. 11d), (5,16,3)→(5,16,2), (6,15,3)→(6,15,2), (7,14,3)→(7,14,2);

and fourth (Fig. 11d)→fifth (Fig. 11e), (5,16,2)→(5,16,1), (6,15,2)→(6,15,1), (7,14,2)→(7,14,1). In the change process, only three sites coupled to neighboring sites by strength $t_{i,j} = 0.01$ are changed and these sites are the nearest neighboring to each other at any two adjacent steps. In this sense, we can view this change process continuously. It means that we need only continuously to modulate three coupling strengths at each step in the change process.

Red spheres in Fig. 11 represent the sites occupied by the localized eigenstates. We can find that for the first $8_3$ knot (Fig. 11a), the third structure (Fig. 11c) and the fifth structure (Fig. 11e), the coordinates of sites occupied by the localized eigenstates are exactly those presented in Table 6; but for the second (Fig. 11b) and fourth (Fig. 11d) structures, some sites with coordinates shown in Table 6 are not occupied by localized eigenstates. The change from Fig. 11a to 11e corresponds to the description of changing one crossing in the text around Fig. 6.

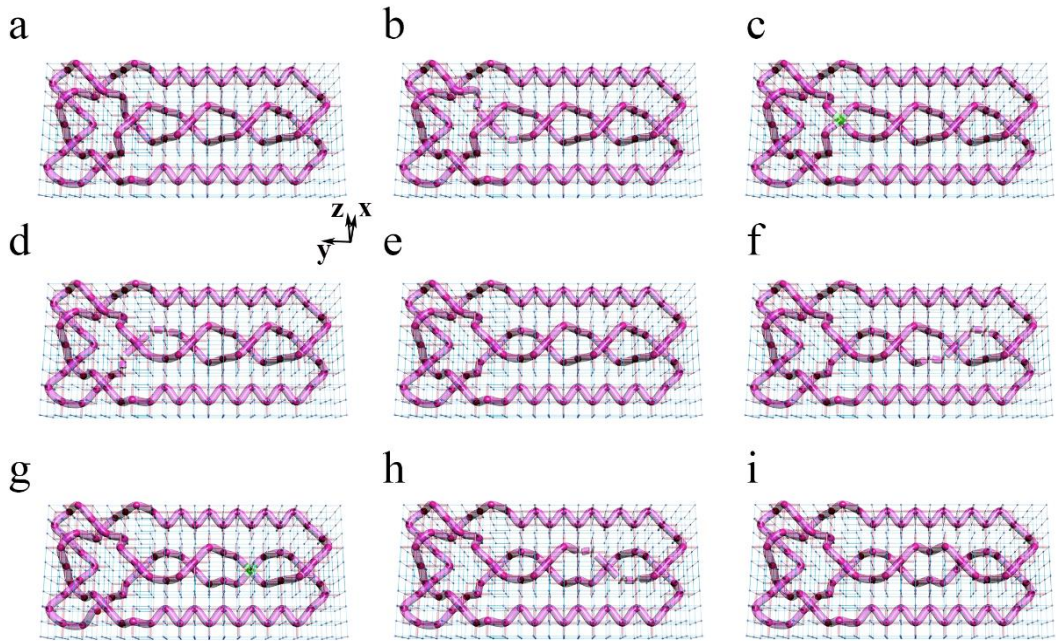

**Fig. 11. Constructions from $8_3$ knot to unknot in the lattices.** From (a) to (i), the $8_3$ knot changes to unknot. Red (blue) cylinders represent the coupling strengths $t = 0.01$ ($s = 1$). (a)-(e) the first change of the crossing emerges in the $8_3$ knot. (e)-(i) the second change of the crossing emerges in the $8_3$ knot. Red spheres in the lattices are the sites occupied by the localized eigenstates. These red spheres are connected in purple tubes.

Then we show the coordinates in Table 7 that connect with the six adjacent sites through the coupling strength $t_{i,j} = 0.01$ for the remaining four structures in Fig. 11 (Fig. 11f, 11g, 11h and 11i), and the coupling strengths between other sites are $s_{k,l} = 1$.

**Table 7**. The coordinates $(x, y, z)$ connect with all adjacent sites through $t_{i,j} = 0.01$

| The sixth structure in Fig. 11f | | | | | | | | | |
|---|---|---|---|---|---|---|---|---|---|
| (2,4,2) | (3,5,2) | (2,6,2) | (3,7,2) | (2,8,2) | (3,9,2) | (2,10,2) | (3,11,2) | (2,12,2) | (3,13,2) | (2,14,2) |
| (2,15,3) | (2,16,2) | (3,17,2) | (4,18,2) | (4,19,1) | (5,20,1) | (6,20,2) | (7,20,3) | (8,19,3) | (8,18,2) | (9,18,1) |
| (10,17,1) | (10,16,2) | (10,15,3) | (10,14,2) | (9,13,2) | (10,12,2) | (9,11,2) | (10,10,2) | (9,9,2) | (10,8,2) | (9,7,2) |
| (10,6,2) | (9,5,2) | (10,4,2) | (9,3,2) | (8,2,2) | (7,2,3) | (6,3,3) | (5,4,3) | (5,5,2) | (5,6,1) | (6,7,1) |
| (7,8,1) | (7,9,2) | (7,10,3) | (6,11,3) | (5,12,3) | (5,13,2) | (5,14,3) | (6,15,3) | (7,16,3) | (8,16,2) | (8,17,3) |
| (9,18,3) | (10,19,3) | (10,20,2) | (9,21,2) | (8,20,2) | (7,19,2) | (6,18,2) | (5,19,2) | (4,20,2) | (3,21,2) | (2,20,2) |
| (2,19,1) | (3,18,1) | (4,17,1) | (4,16,2) | (5,16,1) | (6,15,1) | (7,14,1) | (7,13,2) | (7,12,1) | (6,11,1) | (5,10,1) |
| (5,9,2) | (5,8,2) | (6,7,2) | (7,6,2) | (7,5,2) | (7,4,1) | (6,3,1) | (5,2,1) | (4,2,2) | (3,3,2) | |

| The seventh structure in Fig. 11g | | | | | | | | | |
|---|---|---|---|---|---|---|---|---|---|
| (2,4,2) | (3,5,2) | (2,6,2) | (3,7,2) | (2,8,2) | (3,9,2) | (2,10,2) | (3,11,2) | (2,12,2) | (3,13,2) | (2,14,2) |
| (2,15,3) | (2,16,2) | (3,17,2) | (4,18,2) | (4,19,1) | (5,20,1) | (6,20,2) | (7,20,3) | (8,19,3) | (8,18,2) | (9,18,1) |
| (10,17,1) | (10,16,2) | (10,15,3) | (10,14,2) | (9,13,2) | (10,12,2) | (9,11,2) | (10,10,2) | (9,9,2) | (10,8,2) | (9,7,2) |
| (10,6,2) | (9,5,2) | (10,4,2) | (9,3,2) | (8,2,2) | (7,2,3) | (6,3,3) | (5,4,3) | (5,5,2) | (5,6,1) | (6,7,1) |
| (7,8,1) | (7,9,2) | (7,10,3) | (6,11,3) | (5,12,3) | (5,13,2) | (5,14,3) | (6,15,3) | (7,16,3) | (8,16,2) | (8,17,3) |
| (9,18,3) | (10,19,3) | (10,20,2) | (9,21,2) | (8,20,2) | (7,19,2) | (6,18,2) | (5,19,2) | (4,20,2) | (3,21,2) | (2,20,2) |
| (2,19,1) | (3,18,1) | (4,17,1) | (4,16,2) | (5,16,1) | (6,15,1) | (7,14,1) | (7,13,2) | (7,12,1) | (6,11,1) | (5,10,1) |
| (5,9,2) | (5,8,1) | (6,7,1) | (7,6,1) | (7,5,2) | (7,4,1) | (6,3,1) | (5,2,1) | (4,2,2) | (3,3,2) | |

| The eighth structure in Fig. 11h | | | | | | | | | |
|---|---|---|---|---|---|---|---|---|---|
| (2,4,2) | (3,5,2) | (2,6,2) | (3,7,2) | (2,8,2) | (3,9,2) | (2,10,2) | (3,11,2) | (2,12,2) | (3,13,2) | (2,14,2) |
| (2,15,3) | (2,16,2) | (3,17,2) | (4,18,2) | (4,19,1) | (5,20,1) | (6,20,2) | (7,20,3) | (8,19,3) | (8,18,2) | (9,18,1) |
| (10,17,1) | (10,16,2) | (10,15,3) | (10,14,2) | (9,13,2) | (10,12,2) | (9,11,2) | (10,10,2) | (9,9,2) | (10,8,2) | (9,7,2) |
| (10,6,2) | (9,5,2) | (10,4,2) | (9,3,2) | (8,2,2) | (7,2,3) | (6,3,3) | (5,4,3) | (5,5,2) | (5,6,2) | (6,7,2) |
| (7,8,2) | (7,9,2) | (7,10,3) | (6,11,3) | (5,12,3) | (5,13,2) | (5,14,3) | (6,15,3) | (7,16,3) | (8,16,2) | (8,17,3) |

| (9,18,3) | (10,19,3) | (10,20,2) | (9,21,2) | (8,20,2) | (7,19,2) | (6,18,2) | (5,19,2) | (4,20,2) | (3,21,2) | (2,20,2) |
| --- | --- | --- | --- | --- | --- | --- | --- | --- | --- | --- |
| (2,19,1) | (3,18,1) | (4,17,1) | (4,16,2) | (5,16,1) | (6,15,1) | (7,14,1) | (7,13,2) | (7,12,1) | (6,11,1) | (5,10,1) |
| (5,9,2) | (5,8,1) | (6,7,1) | (7,6,1) | (7,5,2) | (7,4,1) | (6,3,1) | (5,2,1) | (4,2,2) | (3,3,2) | |
| The ninth structure in Fig. 11i | | | | | | | | | | |
| (2,4,2) | (3,5,2) | (2,6,2) | (3,7,2) | (2,8,2) | (3,9,2) | (2,10,2) | (3,11,2) | (2,12,2) | (3,13,2) | (2,14,2) |
| (2,15,3) | (2,16,2) | (3,17,2) | (4,18,2) | (4,19,1) | (5,20,1) | (6,20,2) | (7,20,3) | (8,19,3) | (8,18,2) | (9,18,1) |
| (10,17,1) | (10,16,2) | (10,15,3) | (10,14,2) | (9,13,2) | (10,12,2) | (9,11,2) | (10,10,2) | (9,9,2) | (10,8,2) | (9,7,2) |
| (10,6,2) | (9,5,2) | (10,4,2) | (9,3,2) | (8,2,2) | (7,2,3) | (6,3,3) | (5,4,3) | (5,5,2) | (5,6,3) | (6,7,3) |
| (7,8,3) | (7,9,2) | (7,10,3) | (6,11,3) | (5,12,3) | (5,13,2) | (5,14,3) | (6,15,3) | (7,16,3) | (8,16,2) | (8,17,3) |
| (9,18,3) | (10,19,3) | (10,20,2) | (9,21,2) | (8,20,2) | (7,19,2) | (6,18,2) | (5,19,2) | (4,20,2) | (3,21,2) | (2,20,2) |
| (2,19,1) | (3,18,1) | (4,17,1) | (4,16,2) | (5,16,1) | (6,15,1) | (7,14,1) | (7,13,2) | (7,12,1) | (6,11,1) | (5,10,1) |
| (5,9,2) | (5,8,1) | (6,7,1) | (7,6,1) | (7,5,2) | (7,4,1) | (6,3,1) | (5,2,1) | (4,2,2) | (3,3,2) | |

When compared with the chosen coordinates to realize from the fifth to the ninth structures in the lattices, in Table 7, we find that the coordinates from the fifth to the ninth structures are changed as: fifth (Fig. 11e)→sixth (Fig. 11f), (5,8,3)→(5,8,2), (6,7,3)→(6,7,2), (7,6,3)→(7,6,2); sixth (Fig. 11f)→seventh (Fig. 11g), (5,8,2)→(5,8,1), (6,7,2)→(6,7,1), (7,6,2)→(7,6,1); seventh (Fig. 11g)→eighth (Fig. 11h), (5,6,1)→(5,6,2), (6,7,1)→(6,7,2), (7,8,1)→(7,8,2); and eighth (Fig. 11h)→ninth (Fig. 11i), (5,6,2)→(5,6,3), (6,7,2)→(6,7,3), (7,8,2)→(7,8,3). In the change process, only three sites coupled to neighboring sites by strength $t_{i,j}=0.01$ are changed and these sites are the nearest neighboring to each other at any two adjacent steps. In this sense, we can view this change process continuously. It means that we need only continuously to modulate three coupling strengths at each step in the change process. We can find that for the seventh structure (Fig. 11g) and the ninth structure (Fig. 11i), the coordinates of sites occupied by the localized eigenstates are exactly those presented in Table 7; but for the sixth (Fig. 11f) and eighth (Fig. 11h) structures, some sites with coordinates shown in Table 7 are not occupied by localized eigenstates. The change from Fig. 11e to 11i corresponds to the description of changing one crossing in the text around Fig. 6. Considering the change process making the $8_3$ knot to unknot, we have gone through the change of crossings twice. So the unknotting number for the $8_3$ knot is two.

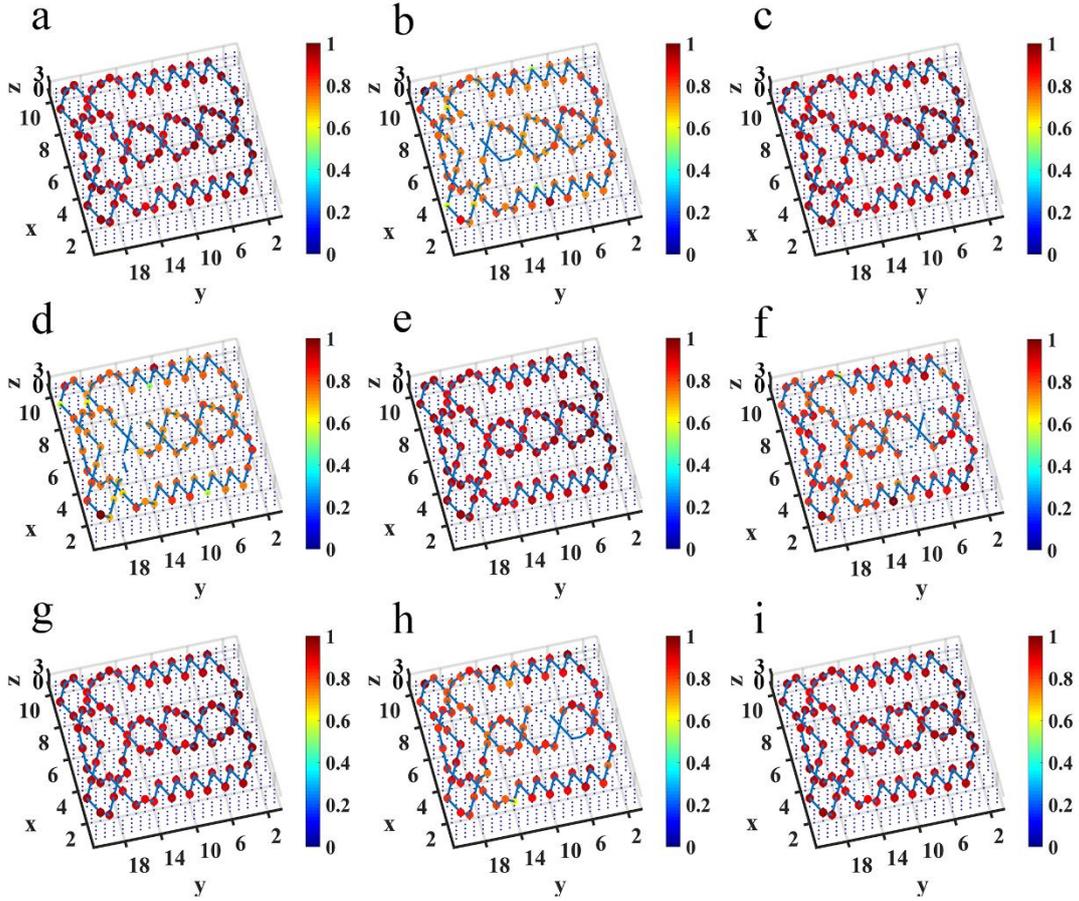

**Fig. 12. Simulated distributions of impedance for the nine structures from $8_3$ knot to unknot.** In all panels, the value of impedances at each node has been normalized to the maximum impedance. From (a) to (i), the distributions of impedance change from $8_3$ knot to unknot.

Similar to the electric realization above, we connect every two nearest neighboring nodes in the electric circuit through the capacitors, and the nodes with the coordinates presented in Table 6 and 7 are connected by small capacitors $C_1$=100pF. Other nodes are connected by large capacitors $C_2$=10nF. Consider the correspondence between the lattice and the electric circuit, we only need to change some capacitors and their associated grounding inductors at each step. In Fig. 12, we provide the distributions of impedances. We can find that for the first five structures, the coordinates of nodes possessing large impedance in the $8_3$ knot, the third structure and the fifth unknot are exactly those presented in Table 6; but in the second and fourth structures, some sites with coordinates shown in Table 6 do not have large impedances. This phenomenon corresponds to the description of changing one crossing in the text around Fig. 6. For the remaining four structures, the coordinates of nodes possessing large impedance in the seventh structure and the ninth unknot are exactly those presented in Table 7; but in the sixth and eighth structures, some sites with coordinates shown in Table 7 do not have large impedances. This phenomenon also corresponds to the description of changing one crossing in the text around Fig. 6. So we need to change the crossings twice to make the $8_3$ knot become unknot.

**Appendix E: The construction details to create electrical knots and links under the action of topoisomerase.**

**a. The change from an unknot to one hopf link**

Here, we describe how to construct knots and links during the action of topoisomerase on the DNA molecules. In Fig. 3a and 3d of the main text, the DNA molecules display the structures of unknot and hopf link. Here, we provide the expression for the structure in Fig. 3a of the main text as

$$f_{deformed-enz} =$$
$$-1 + a_1 x + a_2 y + a_3 z + a_4 x^2 + a_5 xy + a_6 xz + a_7 y^2 + a_8 yz + a_9 z^2 + a_{10} x^3$$
$$+ a_{11} x^2 y + a_{12} x^2 z + a_{13} xy^2 + a_{14} xz^2 + a_{15} xyz + a_{16} y^3 + a_{17} y^2 z + a_{18} yz^2 + a_{19} z^3 + a_{20} x^4$$
$$+ a_{21} x^3 y + a_{22} x^3 z + a_{23} x^2 y^2 + a_{24} x^2 yz + a_{25} x^2 z^2 + a_{26} xy^3 + a_{27} xy^2 z + a_{28} xyz^2 + a_{29} xz^3 + a_{30} y^4 \quad (E1)$$
$$+ a_{31} y^3 z + a_{32} y^2 z^2 + a_{33} yz^3 + a_{34} z^4 + a_{35} x^6 + a_{36} x^4 y^2 + a_{37} x^2 y^4 + a_{38} y^6 + a_{39} zx^4 + a_{40} zx^2 y^2$$
$$+ a_{41} zy^4 + a_{42} z^2 x^4 + a_{43} x^2 y^2 z^2 + a_{44} z^2 y^4 + a_{45} z^3 x^2 + a_{46} z^3 y^2 + a_{47} z^4 x^2 + a_{48} z^4 y^2 + a_{49} z^5$$
$$+ a_{50} z^6 + a_{51} x^5 + a_{52} y^5$$

The structure in Fig. 3a is composed of four fitting curves when we set $|f_{deformed-enz}| \leq 1*10^{-7}$. The ranges of coordinates are $x \in [2,8]$, $y \in [2,16]$ and $z \in [1,3]$. The values of $a_1$ to $a_{52}$ for these four fitting curves are listed below in Table 8 and 9,

**Table 8.** The coefficients $a_1$ to $a_{52}$ in the $f_{deformed-enz}$. The values outside and inside [*] are the coefficients of the first and second fitting curves, respectively.

| | |
|---|---|
| $a_1$ = 0.0366-0.0001i [-8.7080-0.3527i] | $a_2$ =0.4897+0.0002i [-5.8720-0.1232i] |
| $a_3$ =0.5668-0.0004i [-10.9783+1.0580i] | $a_4$ =0.0011-0.0000i [1.0279+0.1103i] |
| $a_5$ =-0.0136+0.0001i [0.5230-0.0286i] | $a_6$ =-0.2117+0.0001i [-2.4155+0.3235i] |
| $a_7$ =-0.0925-0.0001i [0.3322-0.0014i] | $a_8$ =-0.1305+0.0001i [-0.1337+0.1088i] |
| $a_9$ =-0.0312+0.0001i [5.2207-0.5801i] | $a_{10}$ =0.0022+0.0000i [-0.0003+0.0012i] |
| $a_{11}$ =-0.0023+0.0000i [-0.0596+0.0011i] | $a_{12}$ =0.0269+0.0000i [0.4310-0.0472i] |
| $a_{13}$ =0.0017-0.0000i [0.0320-0.0004i] | $a_{14}$ =0.0176-0.0000i [-0.8800+0.0890i] |
| $a_{15}$ =0.0316-0.0000i [-0.0395+0.0045i] | $a_{16}$ =0.0092+0.0000i [0.0385+0.0009i] |
| $a_{17}$ =0.0128-0.0000i [0.0618-0.0027i] | $a_{18}$ =9.78*10$^{-4}$-1.77*10$^{-5}$i [-0.1664+0.0109i] |
| $a_{19}$ =-0.0021-0.0000i [0.7112-0.1040i] | $a_{20}$ =-5.32*10$^{-4}$-3.91*10$^{-8}$i [-0.0142-0.0001i] |
| $a_{21}$ =4.24*10$^{-4}$-1.29*10$^{-7}$i [0.0223-0.0007i] | $a_{22}$ =-0.0019-0.0000i [0.0269-0.0011i] |
| $a_{23}$ =-1.18*10$^{-4}$+1.67*10$^{-7}$i [-0.0099+0.0000i] | $a_{24}$ =-0.0026-0.0000i [-0.0155-0.0044i] |
| $a_{25}$ =-6.10*10$^{-4}$+1.92*10$^{-6}$i [0.0248+0.0004i] | $a_{26}$ =-1.28*10$^{-6}$+2.36*10$^{-7}$i [-0.0011+0.0002i] |
| $a_{27}$ =-0.0013+0.0000i [0.0165-0.0016i] | $a_{28}$ =-0.0017+0.0000i [0.0352-0.0033i] |
| $a_{29}$ =-2.83*10$^{-4}$+2.53*10$^{-6}$i [-0.0212+0.0006i] | $a_{30}$ =-5.13*10$^{-4}$-4.61*10$^{-7}$i [1.91*10$^{-4}$+1.33*10$^{-5}$i] |
| $a_{31}$ =-7.29*10$^{-4}$-2.52*10$^{-7}$i [0.0046-0.0018i] | $a_{32}$ =2.15*10$^{-4}$+5.36*10$^{-7}$i [-0.0174+0.0030i] |
| $a_{33}$ =6.08*10$^{-4}$+3.04*10$^{-7}$i [-0.0423+0.0058i] | $a_{34}$ =-1.95*10$^{-4}$+5.22*10$^{-9}$i [-0.1442+0.0246i] |
| $a_{35}$ =-9.21*10$^{-7}$-2.53*10$^{-10}$i [2.61*10$^{-5}$-1.21*10$^{-6}$i] | $a_{36}$ =-6.91*10$^{-7}$+2.55*10$^{-10}$i [-4.86*10$^{-5}$+2.31*10$^{-6}$i] |
| $a_{37}$ =2.30*10$^{-8}$-4.41*10$^{-10}$i [1.17*10$^{-5}$-9.81*10$^{-7}$i] | $a_{38}$ =-1.77*10$^{-7}$+3.66*10$^{-11}$i [1.17*10$^{-5}$-1.23*10$^{-7}$i] |
| $a_{39}$ =1.04*10$^{-4}$+4.87*10$^{-8}$i [-0.0020+0.0003i] | $a_{40}$ =1.12*10$^{-4}$+2.94*10$^{-8}$i [-0.0017+0.0004i] |
| $a_{41}$ =1.98*10$^{-5}$+7.43*10$^{-9}$i [-3.44*10$^{-4}$+8.24*10$^{-5}$i] | $a_{42}$ =-2.28*10$^{-6}$+3.71*10$^{-9}$i [-4.78*10$^{-5}$+3.00*10$^{-7}$i] |
| $a_{43}$ =6.77*10$^{-6}$-1.13*10$^{-8}$i [4.52*10$^{-5}$-7.48*10$^{-6}$i] | $a_{44}$ =-1.92*10$^{-7}$-5.90*10$^{-10}$i [2.02*10$^{-5}$-3.70*10$^{-6}$i] |
| $a_{45}$ =5.63*10$^{-5}$-2.93*10$^{-7}$i [0.0039-0.0005i] | $a_{46}$ =-3.05*10$^{-5}$-1.88*10$^{-8}$i [0.0025-0.0003i] |
| $a_{47}$ =-7.42*10$^{-6}$+9.20*10$^{-9}$i [6.30*10$^{-6}$+3.32*10$^{-6}$i] | $a_{48}$ =-5.48*10$^{-9}$+7.95*10$^{-10}$i [-6.25*10$^{-5}$+7.69*10$^{-6}$i] |
| $a_{49}$ =5.98*10$^{-5}$-1.37*10$^{-7}$i [0.0214-0.0038i] | $a_{50}$ =-4.29*10$^{-6}$+1.17*10$^{-8}$i [-0.0015+0.0003i] |
| $a_{51}$ =3.05*10$^{-5}$+3.57*10$^{-9}$i [5.73*10$^{-5}$-1.09*10$^{-6}$i] | $a_{52}$ =1.48*10$^{-5}$+7.54*10$^{-9}$i [-3.43*10$^{-4}$-1.41*10$^{-7}$i] |

**Table 9.** The coefficients from $a_1$ to $a_{52}$ in the $f_{deformed-enz}$. The values outside and inside [*] are the coefficients of the third and fourth fitting curves, respectively.

| | |
|---|---|
| $a_1$ = 4.8993-0.2168i [0.5638+0.0000i] | $a_2$ =0.8923-0.0304i [0.1745-0.0000i] |
| $a_3$ =-3.0015+0.2860i [0.6397-0.0000i] | $a_4$ =-0.4972-0.0208i [-0.1295-0.0000i] |
| $a_5$ =-0.7281+0.0383i [-0.0619-0.0000i] | $a_6$ =-0.2780+0.0531i [-0.3271+0.0000i] |
| $a_7$ =0.0165+0.0052i [0.0411+0.0000i] | $a_8$ =0.4229-0.0311i [-0.1675-0.0000i] |
| $a_9$ =-0.8512+0.0344i [-0.0351-0.0000i] | $a_{10}$ =0.1328-0.0019i [0.0145+0.0000i] |
| $a_{11}$ =-0.0088-0.0032i [0.0037+0.0000i] | $a_{12}$ =0.6662+0.0227i [0.0706-0.0000i] |
| $a_{13}$ =-0.0093+0.0025i [-0.0042+0.0000i] | $a_{14}$ =-0.6444-0.0336i [0.0153+0.0000i] |
| $a_{15}$ =-0.2015+0.0029i [0.0410+0.0000i] | $a_{16}$ =-0.0105+0.0008i [-0.0058-0.0000i] |
| $a_{17}$ =0.0537-0.0098i [-0.0140+0.0000i] | $a_{18}$ =0.0193+0.0375i [0.0429-0.0000i] |
| $a_{19}$ =-0.4823-0.0711i [-0.0215+0.0000i] | $a_{20}$ =-0.0130+0.0012i [-6.30*10$^{-4}$-1.28*10$^{-11}$i] |
| $a_{21}$ =0.0010-0.0002i [5.42*10$^{-6}$+1.23*10$^{-12}$i] | $a_{22}$ =-0.2148+0.0068i [-0.0072+0.0000i] |
| $a_{23}$ =-0.0072-0.0000i [4.00*10$^{-4}$-9.66*10$^{-12}$i] | $a_{24}$ =0.0345-0.0006i [-0.0019-0.0000i] |
| $a_{25}$ =0.1795-0.0200i [-0.0028-0.0000i] | $a_{26}$ =0.0123-0.0002i [4.21*10$^{-6}$+6.45*10$^{-12}$i] |
| $a_{27}$ =0.0205-0.0025i [0.0021-0.0000i] | $a_{28}$ =-0.0373+0.0047i [-0.0049+0.0000i] |
| $a_{29}$ =0.0354+0.0028i [0.0038-0.0000i] | $a_{30}$ =-0.0031-0.0001i [3.64*10$^{-4}$+6.91*10$^{-12}$i] |
| $a_{31}$ =0.0004-0.0014i [0.0028+0.0000i] | $a_{32}$ =0.0070+0.0012i [-0.0035+0.0000i] |
| $a_{33}$ =0.0008-0.0027i [-0.0016-0.0000i] | $a_{34}$ =-0.0543+0.0089i [0.0020+0.0000i] |
| $a_{35}$ =-2.19*10$^{-5}$+9.23*10$^{-7}$i [5.14*10$^{-8}$-2.20*10$^{-14}$i] | $a_{36}$ =-6.65*10$^{-5}$-2.71*10$^{-6}$i [-1.40*10$^{-8}$+4.01*10$^{-15}$i] |
| $a_{37}$ =-6.65*10$^{-5}$+2.71*10$^{-6}$i [1.26*10$^{-9}$-1.39*10$^{-14}$i] | $a_{38}$ =2.22*10$^{-5}$-9.03*10$^{-7}$i [6.90*10$^{-8}$+3.30*10$^{-14}$i] |
| $a_{39}$ =0.0131-0.0012i [2.73*10$^{-4}$-1.03*10$^{-12}$i] | $a_{40}$ =-0.0034+0.0002i [-2.84*10$^{-4}$+4.03*10$^{-12}$i] |
| $a_{41}$ =-8.45*10$^{-4}$+2.57*10$^{-4}$i [-1.75*10$^{-4}$+3.72*10$^{-13}$i] | $a_{42}$ =1.63*10$^{-4}$+1.53*10$^{-4}$i [2.07*10$^{-5}$-4.15*10$^{-14}$i] |
| $a_{43}$ =0.0011-0.0000i [4.21*10$^{-5}$-1.53*10$^{-13}$i] | $a_{44}$ =-9.60*10$^{-5}$-2.49*10$^{-5}$i [2.87*10$^{-6}$-5.52*10$^{-14}$i] |
| $a_{45}$ =-0.0323+0.0035i [-4.00*10$^{-5}$+1.57*10$^{-12}$i] | $a_{46}$ =-3.76*10$^{-4}$-2.47*10$^{-4}$i [2.05*10$^{-4}$+4.76*10$^{-13}$i] |
| $a_{47}$ =0.0050-0.0001i [-4.10*10$^{-5}$+1.22*10$^{-13}$i] | $a_{48}$ =1.26*10$^{-4}$+5.26*10$^{-5}$i [2.09*10$^{-6}$+2.68*10$^{-14}$i] |
| $a_{49}$ =0.0012-0.0000i [4.13*10$^{-5}$-5.89*10$^{-12}$i] | $a_{50}$ =-5.17*10$^{-5}$-2.67*10$^{-5}$i [-1.47*10$^{-5}$+4.70*10$^{-13}$i] |
| $a_{51}$ =-3.80*10$^{-6}$-3.43*10$^{-7}$i [-8.81*10$^{-7}$+8.87*10$^{-13}$i] | $a_{52}$ =5.41*10$^{-7}$-4.57*10$^{-8}$i [-2.22*10$^{-6}$-8.48*10$^{-13}$i] |

The other structure shown in Fig. 3d can be obtained in functions in the same way, we do not provide functions here for the lengthy expressions. Here we show how to construct these structures in the lattices. These two structures in Fig. 3a and 3d of the main text are constructed in the lattices as,

$$H_{enzyme} = \sum_{\langle i,j \rangle} t_{i,j} a_i^\dagger a_j + \sum_{\langle k,l \rangle} s_{k,l} a_k^\dagger a_l + \text{H.C.} \quad (E1)$$

The site in the lattice is often described by the integral coordinate. Since we have provided the function describing the structure in Fig. 3a as $|f_{deformed-enz}| \leq 1*10^{-7}$. The ranges of $x$, $y$, $z$ coordinates in the lattice are $x \in [2,8]$, $y \in [2,16]$, $z \in [1,3]$, and the values of $x$, $y$, $z$ are all integers. Therefore, to construct this structure in the lattice, we seek all integral coordinates satisfying $|f_{deformed-enz}| \leq 1*10^{-7}$. We list all these coordinates in sequence in Table 10 below. In the realization of this structure, the coordinates in Table 10 are those connected with the six adjacent sites through the coupling strength $t_{i,j} = 0.01$, and the coupling strengths between other sites are $s_{k,l} = 1$. Moreover, for the structure in Fig. 3d, we also provide the coordinates that connect with the six adjacent sites through the coupling strength $t_{i,j} = 0.01$ in Table 10, and other sites are connected by $s_{k,l}=1$.

**Table 10.** The coordinates $(x,y,z)$ connect with all adjacent sites through $t_{i,j}=0.01$

| Fig. 3a in the main text | | | | | | | | | |
|---|---|---|---|---|---|---|---|---|---|
| (2,3,2) | (2,4,1) | (3,5,1) | (4,6,1) | (4,7,2) | (4,8,3) | (3,9,3) | (2,10,3) | (2,11,2) | (2,12,1) |
| (3,13,1) | (4,14,1) | (5,14,2) | (6,13,2) | (6,12,1) | (6,11,2) | (7,10,2) | (8,11,2) | (8,12,3) | (8,13,2) |
| (8,14,3) | (7,15,3) | (6,16,3) | (5,16,2) | (4,16,3) | (3,16,2) | (2,15,2) | (2,14,3) | (3,13,3) | (4,12,3) |
| (4,11,2) | (4,10,1) | (3,9,1) | (2,8,1) | (2,7,2) | (2,6,3) | (3,5,3) | (4,4,3) | (5,4,2) | (6,5,2) |
| (6,6,3) | (6,7,2) | (7,8,2) | (8,7,2) | (8,6,1) | (8,5,2) | (8,4,1) | (7,3,1) | (6,2,1) | (5,2,2) |
| (4,2,1) | (3,2,2) | | | | | | | | |
| Fig. 3d in the main text | | | | | | | | | |
| (2,3,2) | (2,4,1) | (3,5,1) | (4,6,1) | (4,7,2) | (4,8,3) | (3,9,3) | (2,10,3) | (2,11,2) | (2,12,1) |
| (3,13,1) | (4,14,1) | (5,14,2) | (6,13,2) | (6,12,1) | (6,11,2) | (6,10,1) | (7,9,1) | (8,8,1) | (8,7,2) |
| (8,6,1) | (8,5,2) | (8,4,1) | (7,3,1) | (6,2,1) | (5,2,2) | (4,2,1) | (3,2,2) | (4,4,3) | (3,5,3) |
| (2,6,3) | (2,7,2) | (2,8,1) | (3,9,1) | (4,10,1) | (4,11,2) | (4,12,3) | (3,13,3) | (2,14,3) | (2,15,2) |
| (3,16,2) | (4,16,3) | (5,16,2) | (6,16,3) | (7,15,3) | (8,14,3) | (8,13,2) | (8,12,3) | (8,11,2) | (8,10,3) |
| (7,9,3) | (6,8,3) | (6,7,2) | (6,6,3) | (6,5,2) | (5,4,2) | | | | |

In Figs. 13a and 13b, we provide the realizations of such unknot (Fig. 3a of the main text) and hopf link (Fig. 3d of the main text) in the lattices. Red (blue) cylinders in the lattices represent the coupling strengths $t_{i,j}=0.01$ ($s_{k,l}=1$). Red spheres in the lattices represent the sites occupied by localized eigenstates. As shown in Fig. 13, these sites are exactly those presented in Table 10. To illustrate the distribution clearly, we use the purple and orange tubes in Fig. 13 to connect these sites. We find that they comprise the unknot and hopf link, respectively. So we realize "discrete" versions of unknot and hopf link.

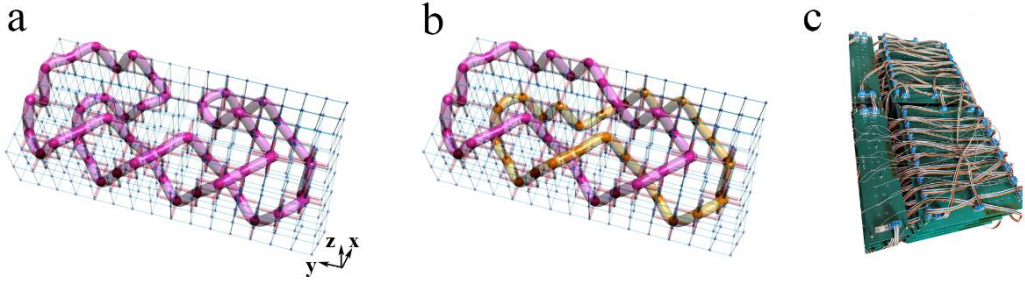

**Fig. 13. Lattices constructions to show the DNA structures of Fig. 3a and Fig. 3d in the main text.** (a) the unknot in Fig. 3a, (b) the hopf link in Fig. 3d. In (a) and (b), red (blue) cylinders represent the coupling strengths $t = 0.01$ ($s = 1$). Red spheres in the lattices are the sites occupied by the localized eigenstates. We connect these sites through red and orange tubes. (c) The electrically experimental setup to realize these two structures.

The corresponding electric circuits are presented in Fig. 13c. Similar to the electric realization above, we connect every two nearest neighboring nodes in the electric circuit through the capacitors. The nodes with the coordinates presented in Table 10 are connected by small capacitors $C_1 = 100$pF, and other nodes are connected by large capacitors $C_2 = 10$nF. Experimentally measured impedances have been shown in Fig. 3 of the main text. Although we provide only one experiment setup in Fig. 13c, we can only modulate some electric capacitors and associated grounding inductors to accomplish the constructions of these two structures in the circuits.

**b. The change from the hopf link to the figure-8 knot**

Actually, the DNA molecule changes to a new topology (figure-8 knot) if the Tn3 resolvase acts on the DNA molecule with the topology in Fig. 3d of the main text again [1]. Here we will show how to realize these two structures in the lattices, the corresponding electric designs can be realized following the way above. These two structures are constructed in the lattices as,

$$H_{enzyme} = \sum_{\langle i,j \rangle} t_{i,j} a_i^\dagger a_j + \sum_{\langle k,l \rangle} s_{k,l} a_k^\dagger a_l + \text{H.C.} \qquad (E2)$$

Here, we do not provide the functions of these two structures for their lengthy expressions. Consider the sites in the lattices are described by the integral coordinates, we provide the coordinates satisfying those functions in Table 11. In the realizations of these two structures, the coordinates in Table 11 connect with the six adjacent sites through the coupling strength $t_{i,j} = 0.01$, and the coupling strengths between other sites are $s_{k,l} = 1$.

**Table 11.** The coordinates $(x, y, z)$ connect with all adjacent sites through $t_{i,j} = 0.01$

| The Tn3 resolvase acts on the DNA molecule once (hopf link) | | | | | | | | | |
|---|---|---|---|---|---|---|---|---|---|
| (2,3,2) | (2,4,1) | (3,5,1) | (4,6,1) | (4,7,2) | (4,8,3) | (3,9,3) | (2,10,3) | (2,11,2) | (2,12,1) | (3,13,1) |
| (4,14,1) | (5,14,2) | (6,13,2) | (6,12,1) | (6,11,2) | (7,10,2) | (8,11,2) | (8,12,3) | (8,13,2) | (9,14,2) | (10,13,2) |
| (10,12,1) | (10,11,2) | (10,10,1) | (11,9,1) | (12,8,1) | (12,7,2) | (12,6,1) | (12,5,2) | (12,4,1) | (11,3,1) | (10,2,1) |
| (9,2,2) | (8,2,1) | (7,2,2) | (6,2,1) | (5,2,2) | (4,2,1) | (3,2,2) | (4,4,3) | (3,5,3) | (2,6,3) | (2,7,2) |
| (2,8,1) | (3,9,1) | (4,10,1) | (4,11,2) | (4,12,3) | (3,13,3) | (2,14,3) | (2,15,2) | (3,16,2) | (4,16,3) | (5,16,2) |
| (6,16,3) | (7,16,2) | (8,16,3) | (9,16,2) | (10,16,3) | (11,15,3) | (12,14,3) | (12,13,2) | (12,12,3) | (12,11,2) | (12,10,3) |
| (11,9,3) | (10,8,3) | (10,7,2) | (10,6,3) | (10,5,2) | (9,4,2) | (8,5,2) | (8,6,1) | (8,7,2) | (7,8,2) | (6,7,2) |
| (6,6,3) | (6,5,2) | (5,4,2) | | | | | | | |
| The Tn3 resolvase acts on the DNA molecule twice (figure-8 knot) | | | | | | | | | |
| (2,3,2) | (2,4,1) | (3,5,1) | (4,6,1) | (4,7,2) | (4,8,3) | (3,9,3) | (2,10,3) | (2,11,2) | (2,12,1) | (3,13,1) |
| (4,14,1) | (5,14,2) | (6,13,2) | (6,12,1) | (6,11,2) | (6,10,1) | (7,9,1) | (8,8,1) | (8,7,2) | (8,6,1) | (8,5,2) |
| (9,4,2) | (10,5,2) | (10,6,3) | (10,7,2) | (10,8,3) | (11,9,3) | (12,10,3) | (12,11,2) | (12,12,3) | (12,13,2) | (12,14,3) |
| (11,15,3) | (10,16,3) | (9,16,2) | (8,16,3) | (7,16,2) | (6,16,3) | (5,16,2) | (4,16,3) | (3,16,2) | (2,15,2) | (2,14,3) |
| (3,13,3) | (4,12,3) | (4,11,2) | (4,10,1) | (3,9,1) | (2,8,1) | (2,7,2) | (2,6,3) | (3,5,3) | (4,4,3) | (5,4,2) |
| (6,5,2) | (6,6,3) | (6,7,2) | (6,8,3) | (7,9,3) | (8,10,3) | (8,11,2) | (8,12,3) | (8,13,2) | (9,14,2) | (10,13,2) |
| (10,12,1) | (10,11,2) | (10,10,1) | (11,9,1) | (12,8,1) | (12,7,2) | (12,6,1) | (12,5,2) | (12,4,1) | (11,3,1) | (10,2,1) |
| (9,2,2) | (8,2,1) | (7,2,2) | (6,2,1) | (5,2,2) | (4,2,1) | (3,2,2) | | | |

In Fig. 14, we provide the realizations of these two structures in the lattices. Red (blue) cylinders in the lattices represent the coupling strengths $t_{i,j} = 0.01$ ($s_{k,l} = 1$).

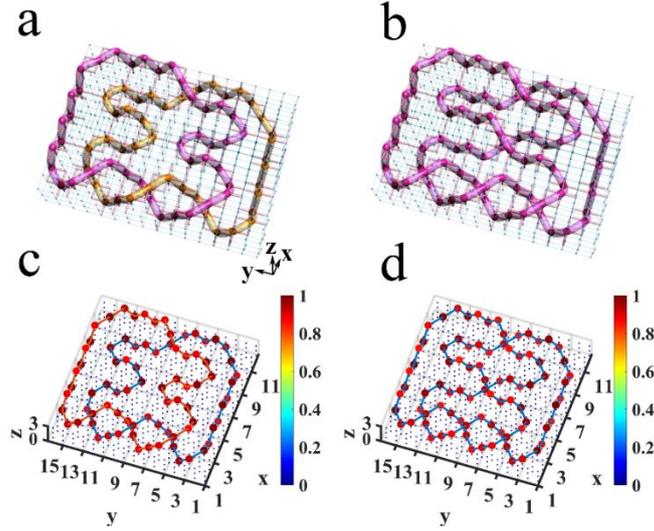

**Fig. 14. Constructions in the lattices and simulated distributions of impedance for these two structures in the electric circuits.** (a) the structure with the Tn3 resolvase acts on the DNA molecule once (hopf link), (b) the structure with the Tn3 resolvase acts on the DNA molecule twice (figure-8 knot). In (a) and (b), red (blue) cylinders represent the coupling strengths $t = 0.01$ ($s = 1$). Red spheres in the lattices are the sites occupied by the localized eigenstates. We connect these sites through red and orange tubes. In (c) and (d), the value of impedance at each node has been normalized to the maximum value.

As shown in Fig. 14, red spheres in the lattices represent the sites occupied by localized eigenstates. These sites are exactly those presented in Table 11. To illustrate the distribution clearly, we use the purple and orange tubes in Fig. 14 to connect these sites. We find that they comprise the "discrete" hopf link and figure-8 knot, respectively. The corresponding electric realizations are similar to the above. We connect every two nearest neighboring nodes in the electric circuit through the capacitors. The nodes with the coordinates presented in Table 11 are connected by small capacitors $C_1$=100pF, and other nodes are connected by large capacitors $C_2$=10nF. The large impedances will emerge at those nodes exactly presented in Table 11.

**Appendix F: Construction details for the DNA structures with different twists and writhes**

**a. Construction details for DNA structures in Figs. 4a, 4d and 4g**

Here, we describe how to construct DNA structures with different twists and writhes. In Figs.

4a, 4d and 4g of the main text, we provide three topologically equivalent structures. The first structure (Fig. 4a) has the twist $Tw = 0$ and the writhe $Wr = 1$. The third structure (Fig. 4g) has the twist $Tw = 1$ and the writhe $Wr = 0$. The electric realizations of these three structures have been provided in Fig. 4b, 4e and 4h, respectively. We provide the expression (Eq. (F1)) for the structure shown in Fig. 4a below,

$$f_{deformed-link} =$$
$$-1 + a_1 x + a_2 y + a_3 z + a_4 x^2 + a_5 xy + a_6 xz + a_7 y^2 + a_8 yz + a_9 z^2 + a_{10} x^3$$
$$+ a_{11} x^2 y + a_{12} x^2 z + a_{13} xy^2 + a_{14} xz^2 + a_{15} xyz + a_{16} y^3 + a_{17} y^2 z + a_{18} yz^2 + a_{19} z^3 + a_{20} x^4$$
$$+ a_{21} x^3 y + a_{22} x^3 z + a_{23} x^2 y^2 + a_{24} x^2 yz + a_{25} x^2 z^2 + a_{26} xy^3 + a_{27} xy^2 z + a_{28} xyz^2 + a_{29} xz^3 + a_{30} y^4 \quad (F1)$$
$$+ a_{31} y^3 z + a_{32} y^2 z^2 + a_{33} yz^3 + a_{34} z^4 + a_{35} x^6 + a_{36} x^4 y^2 + a_{37} x^2 y^4 + a_{38} y^6 + a_{39} zx^4 + a_{40} zx^2 y^2$$
$$+ a_{41} zy^4 + a_{42} z^2 x^4 + a_{43} x^2 y^2 z^2 + a_{44} z^2 y^4 + a_{45} z^3 x^2 + a_{46} z^3 y^2 + a_{47} z^4 x^2 + a_{48} z^4 y^2 + a_{49} z^5$$
$$+ a_{50} z^6 + a_{51} x^5 + a_{52} y^5$$

The structure in Fig. 4a is composed of two fitting curves when we set $|f_{deformed-link}| \leq 1*10^{-8}$. The ranges of coordinates are, $x \in [3,11]$, $y \in [2,14]$, $z \in [1,3]$. The values of $a_1$ to $a_{52}$ for these two fitting curves are listed below in Table 12.

**Table 12.** The coefficients $a_1$ to $a_{52}$ in the $f_{deformed-link}$. The values outside and inside [*] are the coefficients of the first and second fitting curves, respectively.

| | |
|---|---|
| $a_1$ = 0.0632-0.0092i [0.3092-0.0595i] | $a_2$ =0.3050-0.0621i [0.0113-0.0270i] |
| $a_3$ =0.7021+0.0642i [1.0384+0.0997i] | $a_4$ =0.0099+0.0083i [-0.0092+0.0238i] |
| $a_5$ =-0.0083+0.0029i [-0.0285-0.0029i] | $a_6$ =-0.1867+0.0176i [-0.1637+0.0024i] |
| $a_7$ =-0.0285+0.0062i [0.0142+0.0015i] | $a_8$ =-0.1283+0.0421i [-0.1801+0.0755i] |
| $a_9$ =0.1187-0.0650i [0.0505-0.1290i] | $a_{10}$ =0.0012-0.0004i [-0.0041-0.0012i] |
| $a_{11}$ =0.0029-0.0005i [0.0064-0.0012i] | $a_{12}$ =0.0077-0.0079i [0.0003-0.0087i] |
| $a_{13}$ =0.0017+0.0002i [-0.0040+0.0013i] | $a_{14}$ =0.0139+0.0070i [0.0086-0.0040i] |
| $a_{15}$ =0.0142-0.0087i [0.0115-0.0101i] | $a_{16}$ =-0.0008+0.0011i [0.0018+0.0011i] |
| $a_{17}$ =0.0106-0.0085i [0.0118-0.0091i] | $a_{18}$ =-0.0263+0.0207i [-0.0143+0.0193i] |
| $a_{19}$ =-0.0631+0.0161i [-0.1054+0.0236i] | $a_{20}$ =-7.41*10$^{-5}$-6.92*10$^{-5}$i [4.21*10$^{-4}$-1.15*10$^{-4}$i] |
| $a_{21}$ =1.35*10$^{-4}$+5.29*10$^{-5}$i [-4.22*10$^{-4}$+4.06*10$^{-5}$i] | $a_{22}$ =1.05*10$^{-4}$-1.20*10$^{-4}$i [0.0011+0.0001i] |
| $a_{23}$ =2.21*10$^{-5}$+2.17*10$^{-5}$i [-1.82*10$^{-4}$+6.95*10$^{-5}$i] | $a_{24}$ =1.19*10$^{-4}$+1.75*10$^{-4}$i [2.52*10$^{-4}$+9.47*10$^{-4}$i] |
| $a_{25}$ =-0.0018+0.0005i [-0.0012+0.0002i] | $a_{26}$ =-1.35*10$^{-4}$-5.29*10$^{-5}$i [4.22*10$^{-4}$-4.06*10$^{-5}$i] |
| $a_{27}$ =-3.93*10$^{-4}$+6.74*10$^{-4}$i [-0.0015+0.0002i] | $a_{28}$ =0.0011+0.0023i [0.0093+0.0046i] |
| $a_{29}$ =0.0117-0.0030i [0.0007+0.0033i] | $a_{30}$ =1.56*10$^{-4}$-9.24*10$^{-5}$i [-2.37*10$^{-4}$-9.53*10$^{-5}$i] |
| $a_{31}$ =4.28*10$^{-4}$-8.72*10$^{-4}$i [0.0001-0.0012i] | $a_{32}$ =-0.0024+0.0011i [-0.0052+0.0013i] |
| $a_{33}$ =0.0110-0.0073i [0.0282-0.0100i] | $a_{34}$ =0.0002+0.0010i [0.0149+0.0050i] |
| $a_{35}$ =1.61*10$^{-7}$+1.38*10$^{-8}$i [-3.02*10$^{-7}$+1.49*10$^{-8}$i] | $a_{36}$ =-4.83*10$^{-7}$-4.15*10$^{-8}$i [9.08*10$^{-7}$-4.45*10$^{-8}$i] |
| $a_{37}$ =4.83*10$^{-7}$+4.16*10$^{-8}$i [-9.08*10$^{-7}$+4.45*10$^{-8}$i] | $a_{38}$ =-1.61*10$^{-7}$-1.39*10$^{-8}$i [3.02*10$^{-7}$-1.49*10$^{-8}$i] |
| $a_{39}$ =-1.22*10$^{-4}$+1.06*10$^{-4}$i [-1.98*10$^{-4}$+1.51*10$^{-4}$i] | $a_{40}$ =3.87*10$^{-5}$-2.04*10$^{-5}$i [2.64*10$^{-4}$-9.22*10$^{-5}$i] |
| $a_{41}$ =-5.52*10$^{-5}$+1.01*10$^{-4}$i [-6.71*10$^{-5}$+1.28*10$^{-4}$i] | $a_{42}$ =4.30*10$^{-5}$-2.25*10$^{-5}$i [7.05*10$^{-5}$-2.08*10$^{-5}$i] |
| $a_{43}$ =-2.10*10$^{-5}$-1.22*10$^{-5}$i [-1.17*10$^{-4}$-1.54*10$^{-5}$i] | $a_{44}$ =1.27*10$^{-5}$-1.19*10$^{-5}$i [4.64*10$^{-5}$-1.06*10$^{-5}$i] |
| $a_{45}$ =-0.0011+0.0001i [-0.0013-0.0009i] | $a_{46}$ =1.64*10$^{-4}$-6.35*10$^{-5}$i [-0.0013+0.0002i] |
| $a_{47}$ =-8.66*10$^{-5}$+4.50*10$^{-5}$i [-1.40*10$^{-4}$+4.15*10$^{-5}$i] | $a_{48}$ =-1.55*10$^{-4}$+7.62*10$^{-5}$i [-2.22*10$^{-4}$-1.38*10$^{-5}$i] |
| $a_{49}$ =0.0013-0.0003i [0.0015+0.0008i] | $a_{50}$ =4.33*10$^{-5}$-2.25*10$^{-5}$i [6.98*10$^{-5}$-2.08*10$^{-5}$i] |
| $a_{51}$ =4.24*10$^{-10}$-3.53*10$^{-11}$i [-8.44*10$^{-9}$-1.73*10$^{-9}$i] | $a_{52}$ =-1.43*10$^{-9}$+5.37*10$^{-10}$i [6.21*10$^{-9}$+1.35*10$^{-9}$i] |

Other structures shown in Fig. 4d and Fig. 4g can be obtained in functions in the same way, and we do not provide the detailed lengthy expressions here. Then we show how to construct these structures in the lattices. These three structures in Fig. 4 of the main text are constructed in the lattices as,

$$H_{dna} = \sum_{\langle i,j \rangle} t_{i,j} a_i^\dagger a_j + \sum_{\langle k,l \rangle} s_{k,l} a_k^\dagger a_l + \text{H.C.} \quad (F2)$$

The site in the lattice is often described by the integral coordinate. Since we have provided the function describing the structure in Fig. 4a as $|f_{deformed-link}| \leq 1*10^{-8}$. The ranges of $x$, $y$, $z$ coordinates in the lattice are $x \in [3,11]$, $y \in [2,14]$, $z \in [1,3]$, and the values of $x$, $y$, $z$ are all integers. Therefore, to construct this structure in the lattice, we seek all integral coordinates satisfying $|f_{deformed-link}| \leq 1*10^{-8}$. We list all these coordinates in sequence in Table 13 below. In the realization of this structure, these coordinates connecting with the six adjacent sites through the coupling strength $t_{i,j} = 0.01$, and the coupling strengths between other sites are $s_{k,l} = 1$. Moreover, for the structures in Fig. 4d and Fig. 4g, we also provide the coordinates connecting with the six adjacent sites through the coupling strength $t_{i,j} = 0.01$ in Table 13.

**Table 13.** The coordinates $(x, y, z)$ connect with all adjacent sites through $t_{i,j} = 0.01$

| The structure in Fig. 4a of the main text | | | | | | | | | |
|---|---|---|---|---|---|---|---|---|---|
| (3,5,1) | (4,6,1) | (5,7,1) | (6,8,1) | (7,9,1) | (8,10,1) | (9,11,1) | (8,12,1) | (7,12,2) | (6,12,3) | (5,11,3) |
| (6,10,3) | (7,9,3) | (8,8,3) | (9,7,3) | (10,6,3) | (11,5,3) | (10,4,3) | (9,3,3) | (8,2,3) | (7,2,2) | (6,2,1) |
| (5,3,1) | (4,4,1) | (5,5,1) | (6,6,1) | (7,7,1) | (8,8,1) | (9,9,1) | (10,10,1) | (11,11,1) | (10,12,1) | (9,13,1) |
| (8,14,1) | (7,14,2) | (6,14,3) | (5,13,3) | (4,12,3) | (3,11,3) | (4,10,3) | (5,9,3) | (6,8,3) | (7,7,3) | (8,6,3) |
| (9,5,3) | (8,4,3) | (7,4,2) | (6,4,1) | | | | | | | |

| The structure in Fig. 4d of the main text | | | | | | | | | |
|---|---|---|---|---|---|---|---|---|---|
| (2,6,2) | (2,7,1) | (2,8,2) | (2,9,1) | (2,10,2) | (3,11,2) | (4,11,1) | (5,11,2) | (6,11,1) | (7,11,2) | (7,12,3) |
| (6,12,4) | (7,11,4) | (8,11,5) | (9,11,4) | (10,11,5) | (11,11,4) | (12,10,4) | (12,9,5) | (12,8,4) | (12,7,5) | (12,6,4) |
| (11,5,4) | (10,4,4) | (9,3,4) | (8,2,4) | (7,2,3) | (6,2,2) | (5,3,2) | (4,4,2) | (3,5,2) | (4,6,2) | (4,7,1) |
| (4,8,2) | (4,9,1) | (5,9,2) | (6,9,1) | (7,9,2) | (8,9,1) | (9,10,1) | (10,11,1) | (10,12,2) | (9,13,2) | (8,14,2) |
| (7,14,3) | (6,14,4) | (5,13,4) | (4,12,4) | (4,11,5) | (4,10,4) | (5,9,4) | (6,9,5) | (7,9,4) | (8,9,5) | (9,9,4) |
| (10,9,5) | (10,8,4) | (10,7,5) | (10,6,4) | (9,5,4) | (8,4,4) | (7,4,3) | (6,4,2) | (5,5,2) | | |

| The structure in Fig. 4g of the main text | | | | | | | | | |
|---|---|---|---|---|---|---|---|---|---|
| (2,6,2) | (2,7,1) | (2,8,2) | (2,9,1) | (2,10,2) | (3,11,2) | (4,11,1) | (5,11,2) | (6,11,1) | (7,11,2) | (7,12,3) |
| (8,12,4) | (9,11,4) | (10,10,4) | (11,10,3) | (12,9,3) | (12,8,4) | (12,7,5) | (12,6,4) | (11,5,4) | (10,4,4) | (9,3,4) |
| (8,2,4) | (7,2,3) | (6,2,2) | (5,3,2) | (4,4,2) | (3,5,2) | (4,6,2) | (4,7,1) | (4,8,2) | (4,9,1) | (5,9,2) |

| (6,9,1) | (7,9,2) | (8,9,1) | (9,10,1) | (10,11,1) | (10,12,2) | (9,13,2) | (8,13,1) | (8,14,2) | (9,14,3) | (10,14,4) |
|---|---|---|---|---|---|---|---|---|---|---|
| (11,14,5) | (12,13,5) | (12,12,4) | (12,11,5) | (11,10,5) | (10,9,5) | (10,8,4) | (10,7,5) | (10,6,4) | (9,5,4) | (8,4,4) |
| (7,4,3) | (6,4,2) | (5,5,2) | | | | | | | | |

In Figs. 15a-15c, we provide the realizations of these three structures in the lattices. Red (blue) cylinders in the lattices represent the coupling strengths $t_{i,j}=0.01$ ($s_{k,l}=1$). Red spheres in the lattices represent the sites occupied by localized eigenstates. As shown in Figs. 15a-15c, these sites are exactly those presented in Table 13. To illustrate the distribution clearly, we use the purple and orange tubes in Figs. 15a-15c to connect these sites. We find that they comprise the first structure with the twist $Tw=0$ and the writhe $Wr=1$ (Fig. 15a), the second structure (Fig. 15b) and the third structure with the twist $Tw=1$ and the writhe $Wr=0$ (Fig. 15c), respectively. So we realize "discrete" versions of three structures in Figs. 4a, 4d and 4g of the main text.

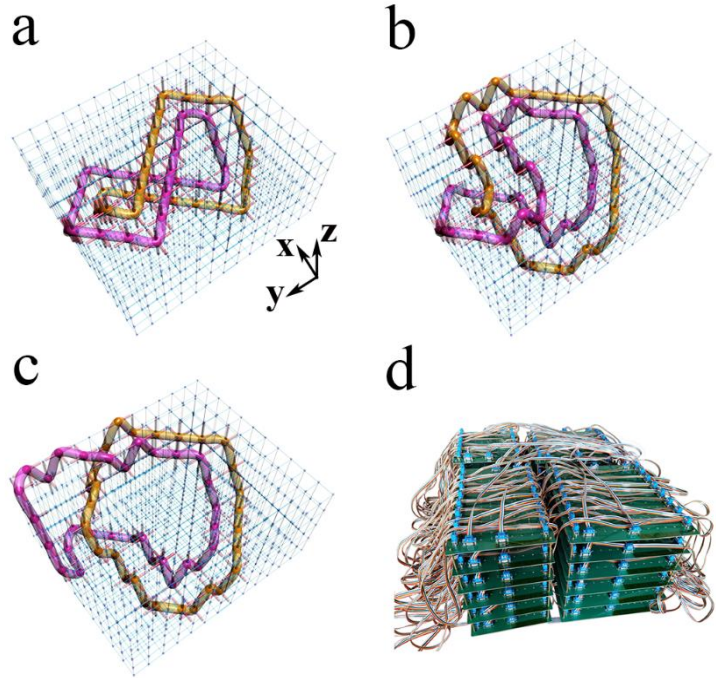

**Fig. 15. Constructions of DNA structures with three different geometries in the lattices.** (a) the twist $Tw=0$ and the writhe $Wr=1$, same as that presented in Fig. 4a of the main text. (b) same as in Fig. 4d of the main text. (c) the twist $Tw=1$ and the writhe $Wr=0$, same as in Fig. 4g of the main text. In (a)-(c), red (blue) cylinders represent the coupling strengths

$t = 0.01$ ($s = 1$). Red spheres denote the sites occupied by localized eigenstates. We connect these sites in one purple and orange tubes. (d) The corresponding electrically experimental setup to realize these three structures.

The corresponding electric circuits are also presented in Fig. 15d. Similar to the electric realization above, we connect every two nodes in the electric circuit through the capacitors. The nodes with the coordinates presented in Table 13 are connected by $C_1$=100pF, and other nodes are connected by $C_2$=10nF. The corresponding experimental designs have been provided in Figs. 4b, 4e and 4h of the main text, respectively. Experimentally measured impedances in these circuits have been shown in Figs. 4c, 4f and 4i of the main text, respectively.

**b. Construction details of DNA structures in the middle of change process**

To demonstrate these two structures ($\text{Tw} = 0$ and $\text{Wr} = 1$, $\text{Tw} = 1$ and $\text{Wr} = 0$) are topological equivalent, we need to start from the structure with $\text{Tw} = 0$ and $\text{Wr} = 1$, and finally reach the structure with $\text{Tw} = 1$ and $\text{Wr} = 0$ continuously. Constructions of these structures are similar to the description in Appendix A. That is shown as

$$H_{dna} = \sum_{\langle i,j \rangle} t_{i,j} a_i^\dagger a_j + \sum_{\langle k,l \rangle} s_{k,l} a_k^\dagger a_l + \text{H.C.} \tag{F3}$$

During this change process, we need to modulate the corresponding designs continuously, and do not change the crossing at any structures. In the change process, only some sites coupled to neighboring sites by the strength $t_{i,j} = 0.01$ are changed and these sites are the nearest neighboring to each other at any two adjacent steps. In this sense, we can view this change process continuously. In the following, we provide the details of remaining structures in this continuous change process in the lattices. In the realizations of different structures between the first structure in Fig. 4a and the second structure in Fig. 4d, the coordinates in Table 14 connect with the six adjacent sites through the coupling strength $t_{i,j} = 0.01$, and the coupling strengths between other sites are $s_{k,l} = 1$.

**Table 14.** The coordinates $(x, y, z)$ connect with all adjacent sites through $t_{i,j} = 0.01$

| The structure in Fig. 16a | | | | | | | | | |
|---|---|---|---|---|---|---|---|---|---|
| (3,5,2) | (3,6,1) | (4,7,1) | (5,8,1) | (6,9,1) | (7,10,1) | (8,10,2) | (9,11,2) | (8,12,2) | (7,12,3) |
| (6,12,4) | (5,11,4) | (6,10,4) | (7,10,5) | (8,9,5) | (9,8,5) | (10,7,5) | (11,6,5) | (11,5,4) | (10,4,4) |
| (9,3,4) | (8,2,4) | (7,2,3) | (6,2,2) | (5,3,2) | (4,4,2) | (5,5,2) | (5,6,1) | (6,7,1) | (7,8,1) |
| (8,9,1) | (9,9,2) | (10,10,2) | (11,11,2) | (10,12,2) | (9,13,2) | (8,14,2) | (7,14,3) | (6,14,4) | (5,13,4) |
| (4,12,4) | (3,11,4) | (4,10,4) | (5,9,4) | (6,9,5) | (7,8,5) | (8,7,5) | (9,6,5) | (9,5,4) | (8,4,4) |
| (7,4,3) | (6,4,2) | | | | | | | | |
| The structure in Fig. 16c | | | | | | | | | |
| (2,6,2) | (3,7,2) | (4,8,2) | (5,9,2) | (6,10,2) | (7,10,1) | (8,10,2) | (9,11,2) | (8,12,2) | (7,12,3) |
| (6,12,4) | (5,11,4) | (6,10,4) | (7,10,5) | (8,10,4) | (9,9,4) | (10,8,4) | (11,7,4) | (12,6,4) | (11,5,4) |
| (10,4,4) | (9,3,4) | (8,2,4) | (7,2,3) | (6,2,2) | (5,3,2) | (4,4,2) | (3,5,2) | (4,6,2) | (5,7,2) |
| (6,8,2) | (7,8,1) | (8,9,1) | (9,9,2) | (10,10,2) | (11,11,2) | (10,12,2) | (9,13,2) | (8,14,2) | (7,14,3) |
| (6,14,4) | (5,13,4) | (4,12,4) | (3,11,4) | (4,10,4) | (5,9,4) | (6,9,5) | (7,9,4) | (8,8,4) | (9,7,4) |
| (10,6,4) | (9,5,4) | (8,4,4) | (7,4,3) | (6,4,2) | (5,5,2) | | | | |
| The structure in Fig. 16e | | | | | | | | | |
| (2,6,2) | (2,7,1) | (3,8,1) | (4,9,1) | (5,10,1) | (6,11,1) | (7,11,2) | (8,12,2) | (7,12,3) | (6,12,4) |
| (7,11,4) | (8,11,5) | (9,10,5) | (10,9,5) | (11,8,5) | (12,7,5) | (12,6,4) | (11,5,4) | (10,4,4) | (9,3,4) |
| (8,2,4) | (7,2,3) | (6,2,2) | (5,3,2) | (4,4,2) | (3,5,2) | (4,6,2) | (4,7,1) | (5,8,1) | (6,9,1) |
| (7,9,2) | (8,9,1) | (9,9,2) | (10,10,2) | (11,11,2) | (10,12,2) | (9,13,2) | (8,14,2) | (7,14,3) | (6,14,4) |
| (5,13,4) | (4,12,4) | (3,11,4) | (4,10,4) | (5,9,4) | (6,9,5) | (7,9,4) | (8,9,5) | (9,8,5) | (10,7,5) |
| (10,6,4) | (9,5,4) | (8,4,4) | (7,4,3) | (6,4,2) | (5,5,2) | | | | |
| The structure in Fig. 16g | | | | | | | | | |
| (2,6,2) | (2,7,1) | (2,8,2) | (3,9,2) | (4,10,2) | (5,11,2) | (6,11,1) | (7,11,2) | (8,12,2) | (7,12,3) |
| (6,12,4) | (7,11,4) | (8,11,5) | (9,11,4) | (10,10,4) | (11,9,4) | (12,8,4) | (12,7,5) | (12,6,4) | (11,5,4) |
| (10,4,4) | (9,3,4) | (8,2,4) | (7,2,3) | (6,2,2) | (5,3,2) | (4,4,2) | (3,5,2) | (4,6,2) | (4,7,1) |
| (4,8,2) | (5,9,2) | (6,9,1) | (7,9,2) | (8,9,1) | (9,10,1) | (10,11,1) | (10,12,2) | (9,13,2) | (8,14,2) |
| (7,14,3) | (6,14,4) | (5,13,4) | (4,12,4) | (3,11,4) | (4,10,4) | (5,9,4) | (6,9,5) | (7,9,4) | (8,9,5) |
| (9,8,5) | (10,8,4) | (10,7,5) | (10,6,4) | (9,5,4) | (8,4,4) | (7,4,3) | (6,4,2) | (5,5,2) | |
| The structure in Fig. 16i | | | | | | | | | |

| (2,6,2) | (2,7,1) | (2,8,2) | (2,9,1) | (3,10,1) | (4,11,1) | (5,11,2) | (6,11,1) | (7,11,2) | (8,12,2) |
|---|---|---|---|---|---|---|---|---|---|
| (7,12,3) | (6,12,4) | (7,11,4) | (8,11,5) | (9,11,4) | (10,11,5) | (11,10,5) | (12,9,5) | (12,8,4) | (12,7,5) |
| (12,6,4) | (11,5,4) | (10,4,4) | (9,3,4) | (8,2,4) | (7,2,3) | (6,2,2) | (5,3,2) | (4,4,2) | (3,5,2) |
| (4,6,2) | (4,7,1) | (4,8,2) | (4,9,1) | (5,9,2) | (6,9,1) | (7,9,2) | (8,9,1) | (9,10,1) | (10,11,1) |
| (10,12,2) | (9,13,2) | (8,14,2) | (7,14,3) | (6,14,4) | (5,13,4) | (4,12,4) | (3,11,4) | (4,10,4) | (5,9,4) |
| (6,9,5) | (7,9,4) | (8,9,5) | (9,9,4) | (10,9,5) | (10,8,4) | (10,7,5) | (10,6,4) | (9,5,4) | (8,4,4) |
| (7,4,3) | (6,4,2) | (5,5,2) | | | | | | | |

In Fig. 16, we provide the realizations of these five structures in the lattices.

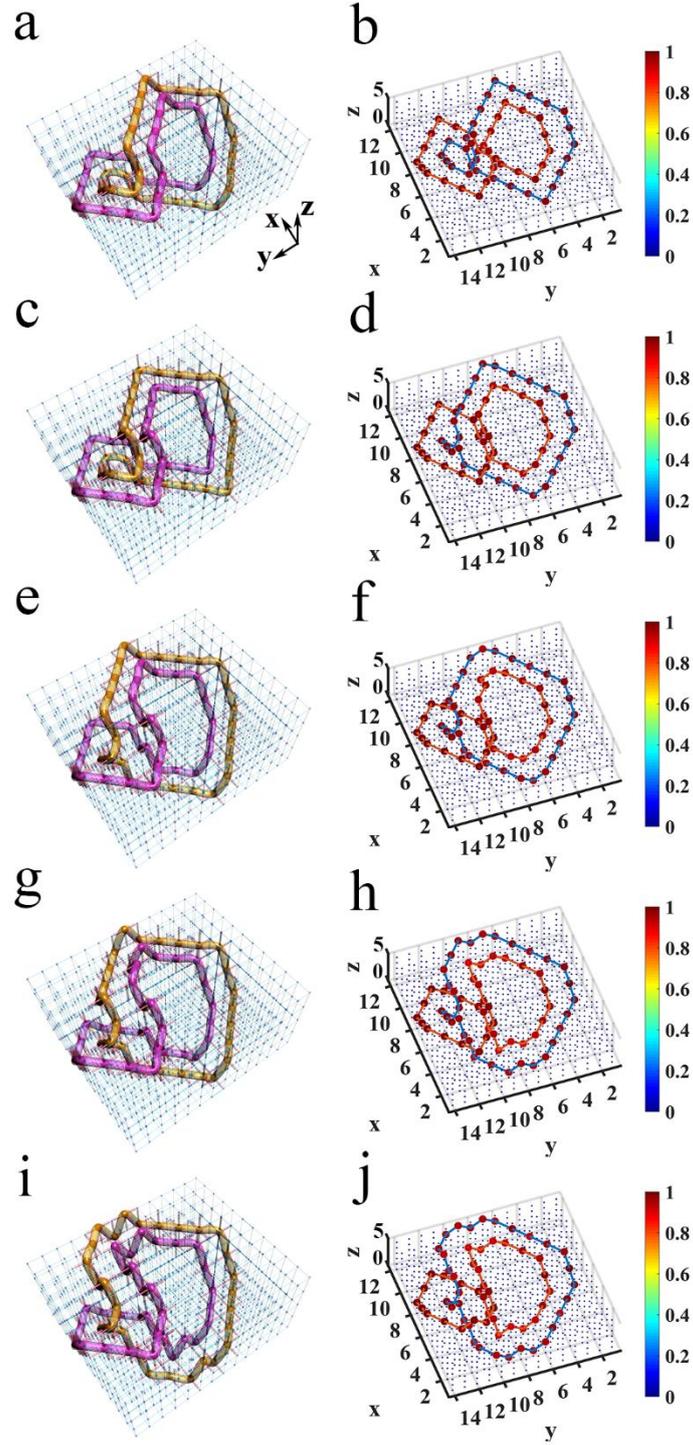

**Fig. 16. Constructions of five structures between the first structure (Fig. 4a) and the second structure (Fig. 4d) in the continuous change process.** In (a), (c), (e), (g) and (i), red (blue) cylinders represent the coupling strengths $t = 0.01$ ($s = 1$) in the lattices. Only some sites coupled to neighboring sites by strength $t_{i,j} = 0.01$ are changed and these sites are nearest neighboring to each other at any two adjacent steps. Red spheres denote the sites occupied by

localized eigenstates. We connect these sites in purple and orange tubes. In (b), (d), (f), (h) and (j), we provide the simulated distributions of impedance for electric realizations of (a), (c), (e), (g) and (i), respectively. The value of impedances at each node have been normalized to the maximum impedance.

Red (blue) cylinders in the lattices represent the coupling strengths $t_{i,j}=0.01$ ($s_{k,l}=1$). Red spheres in the lattices represent the sites occupied by localized eigenstates. As shown in Figs. 16a, 16c, 16e, 16g and 16i, these sites are exactly those presented in Table 14. To illustrate the distribution clearly, we use the purple and orange tubes in Fig. 16 to connect these sites. We find that they comprise the five structures. So we realize "discrete" versions of these five structures. The corresponding electric realizations are similar to those above. We connect every two nodes in the electric circuit through the electric capacitors. The nodes with the coordinates presented in Table 14 are connected by the small capacitor $C_1$=100pF, and other nodes are connected by the large capacitor $C_2$=10nF. The distributions of large impedances are consistent with those of localized eigenstates in Figs. 16b, 16d, 16f, 16h and 16j. We can find that the nodes having large impedances in the circuits are consistent with those sites occupied by localized eigenstates in the lattices.

We also provide the continuous change between the structure in Fig. 4d and the structure in Fig. 4f. To realize such continuous change, we need to construct another five structures. The coordinates in Table 15 connect with the six adjacent sites through the coupling strength $t_{i,j}=0.01$, and the coupling strengths between other sites are $s_{k,l}=1$.

**Table 15.** The coordinates $(x,y,z)$ connect with all adjacent sites through $t_{i,j}=0.01$

| The structure in Fig. 17a | | | | | | | | | |
|---|---|---|---|---|---|---|---|---|---|
| (2,6,2) | (2,7,1) | (2,8,2) | (2,9,1) | (2,10,2) | (3,11,2) | (4,11,1) | (5,11,2) | (6,11,1) | (7,11,2) |
| (7,12,3) | (8,12,4) | (9,11,4) | (10,11,3) | (11,10,3) | (11,9,4) | (12,8,4) | (12,7,5) | (12,6,4) | (11,5,4) |
| (10,4,4) | (9,3,4) | (8,2,4) | (7,2,3) | (6,2,2) | (5,3,2) | (4,4,2) | (3,5,2) | (4,6,2) | (4,7,1) |
| (4,8,2) | (4,9,1) | (5,9,2) | (6,9,1) | (7,9,2) | (8,9,1) | (9,10,1) | (10,11,1) | (10,12,2) | (9,13,2) |
| (8,14,2) | (7,14,3) | (6,14,4) | (5,13,4) | (4,12,4) | (5,11,4) | (6,10,4) | (7,10,5) | (8,10,4) | (9,9,4) |

| (9,8,5) | (10,7,5) | (10,6,4) | (9,5,4) | (8,4,4) | (7,4,3) | (6,4,2) | (5,5,2) | | |
|---|---|---|---|---|---|---|---|---|---|
| The structure in Fig. 17c | | | | | | | | | |
| (2,6,2) | (2,7,1) | (2,8,2) | (2,9,1) | (2,10,2) | (3,11,2) | (4,11,1) | (5,11,2) | (6,11,1) | (7,11,2) |
| (7,12,3) | (8,12,4) | (9,11,4) | (10,11,3) | (11,10,3) | (12,9,3) | (12,8,4) | (12,7,5) | (12,6,4) | (11,5,4) |
| (10,4,4) | (9,3,4) | (8,2,4) | (7,2,3) | (6,2,2) | (5,3,2) | (4,4,2) | (3,5,2) | (4,6,2) | (4,7,1) |
| (4,8,2) | (4,9,1) | (5,9,2) | (6,9,1) | (7,9,2) | (8,9,1) | (9,10,1) | (10,11,1) | (10,12,2) | (9,13,2) |
| (8,14,2) | (7,14,3) | (6,14,4) | (6,13,5) | (5,12,5) | (6,11,5) | (7,10,5) | (8,10,4) | (9,10,5) | (10,9,5) |
| (10,8,4) | (10,7,5) | (10,6,4) | (9,5,4) | (8,4,4) | (7,4,3) | (6,4,2) | (5,5,2) | | |
| The structure in Fig. 17e | | | | | | | | | |
| (2,6,2) | (2,7,1) | (2,8,2) | (2,9,1) | (2,10,2) | (3,11,2) | (4,11,1) | (5,11,2) | (6,11,1) | (7,11,2) |
| (7,12,3) | (8,12,4) | (9,11,4) | (10,10,4) | (11,10,3) | (12,9,3) | (12,8,4) | (12,7,5) | (12,6,4) | (11,5,4) |
| (10,4,4) | (9,3,4) | (8,2,4) | (7,2,3) | (6,2,2) | (5,3,2) | (4,4,2) | (3,5,2) | (4,6,2) | (4,7,1) |
| (4,8,2) | (4,9,1) | (5,9,2) | (6,9,1) | (7,9,2) | (8,9,1) | (9,10,1) | (10,11,1) | (10,12,2) | (9,13,2) |
| (8,14,2) | (7,14,3) | (7,13,4) | (7,12,5) | (8,11,5) | (9,10,5) | (9,9,4) | (10,8,4) | (10,7,5) | (10,6,4) |
| (9,5,4) | (8,4,4) | (7,4,3) | (6,4,2) | (5,5,2) | | | | | |
| The structure in Fig. 17g | | | | | | | | | |
| (2,6,2) | (2,7,1) | (2,8,2) | (2,9,1) | (2,10,2) | (3,11,2) | (4,11,1) | (5,11,2) | (6,11,1) | (7,11,2) |
| (7,12,3) | (8,12,4) | (9,11,4) | (10,10,4) | (11,10,3) | (12,9,3) | (12,8,4) | (12,7,5) | (12,6,4) | (11,5,4) |
| (10,4,4) | (9,3,4) | (8,2,4) | (7,2,3) | (6,2,2) | (5,3,2) | (4,4,2) | (3,5,2) | (4,6,2) | (4,7,1) |
| (4,8,2) | (4,9,1) | (5,9,2) | (6,9,1) | (7,9,2) | (8,9,1) | (9,10,1) | (10,11,1) | (10,12,2) | (9,13,2) |
| (8,14,2) | (7,14,3) | (8,14,4) | (9,13,4) | (9,12,5) | (10,11,5) | (11,10,5) | (10,9,5) | (10,8,4) | (10,7,5) |
| (10,6,4) | (9,5,4) | (8,4,4) | (7,4,3) | (6,4,2) | (5,5,2) | | | | |
| The structure in Fig. 17i | | | | | | | | | |
| (2,6,2) | (2,7,1) | (2,8,2) | (2,9,1) | (2,10,2) | (3,11,2) | (4,11,1) | (5,11,2) | (6,11,1) | (7,11,2) |
| (7,12,3) | (8,12,4) | (9,11,4) | (10,10,4) | (11,10,3) | (12,9,3) | (12,8,4) | (12,7,5) | (12,6,4) | (11,5,4) |
| (10,4,4) | (9,3,4) | (8,2,4) | (7,2,3) | (6,2,2) | (5,3,2) | (4,4,2) | (3,5,2) | (4,6,2) | (4,7,1) |
| (4,8,2) | (4,9,1) | (5,9,2) | (6,9,1) | (7,9,2) | (8,9,1) | (9,10,1) | (10,11,1) | (10,12,2) | (9,13,2) |
| (8,14,2) | (9,14,3) | (10,14,4) | (11,13,4) | (11,12,5) | (12,11,5) | (11,10,5) | (10,9,5) | (10,8,4) | (10,7,5) |
| (10,6,4) | (9,5,4) | (8,4,4) | (7,4,3) | (6,4,2) | (5,5,2) | | | | |

In Fig. 17, we provide the realizations of these five structures in the lattices. Red (blue) cylinders in the lattices represent the coupling strengths $t_{i,j} = 0.01$ ($s_{k,l} = 1$). Red spheres in the lattices represent the sites occupied by localized eigenstates. As shown in Fig. 17a, 17c, 17e, 17g and 17i, these sites are exactly those presented in Table 15. To illustrate the distribution clearly, we use the purple and orange tubes in Fig. 17 to connect these sites. We find that they comprise the five structures. So we realize "discrete" versions of these five structures. The corresponding electric realizations are similar to those above. We connect every two nodes in the electric circuit through the electric capacitors. The nodes with the coordinates presented in Table 15 are connected by the small electric capacitor $C_1 = 100\text{pF}$, and other nodes are connected by the large electric capacitor $C_2 = 10\text{nF}$. The distributions of large impedances are consistent with those of localized eigenstates in Figs. 17b, 17d, 17f, 17h and 17j. We can find that the nodes having large impedances in the circuits are consistent with those sites occupied by localized eigenstates in the lattices.

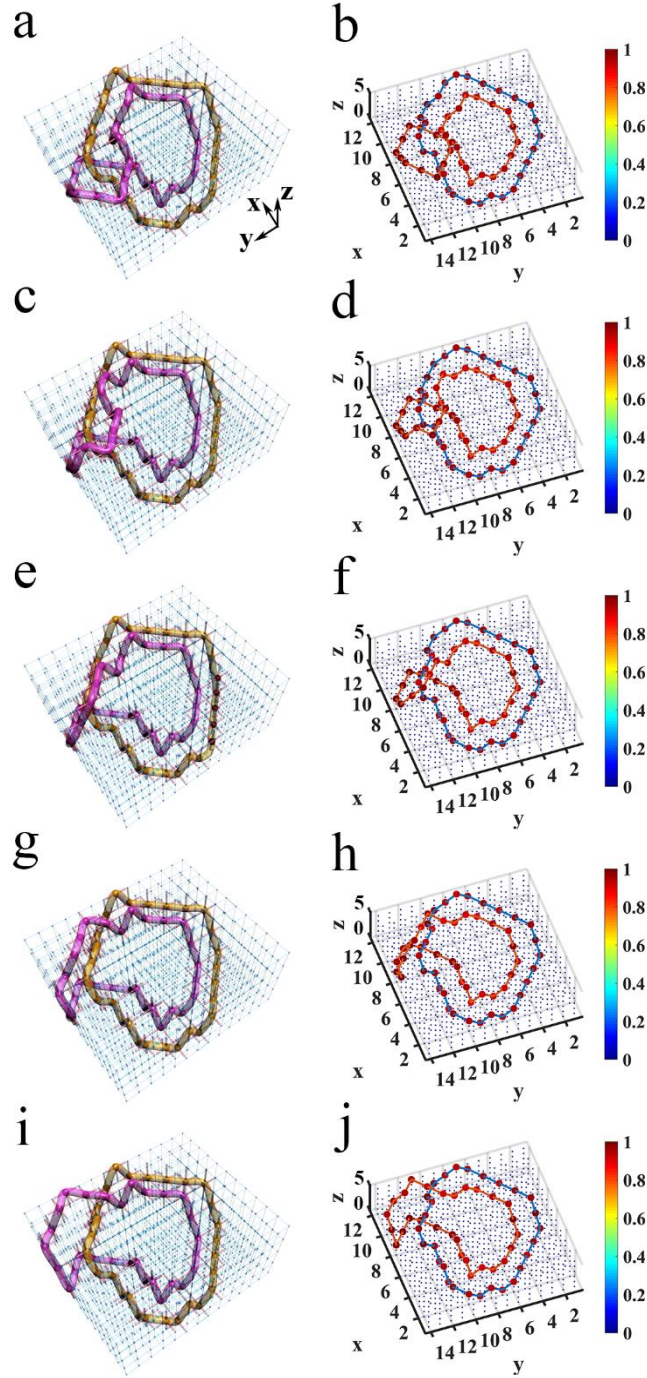

**Fig. 17. Constructions of five structures between the second structure (Fig. 4d) and the third structure (Fig. 4g) in the continuous change process.** In (a), (c), (e), (g) and (i), red (blue) cylinders represent the coupling strengths $t = 0.01$ ($s = 1$) in the lattices. Red spheres denote the sites occupied by localized eigenstates. We connect these sites in purple and orange tubes. In (b), (d), (f), (h) and (j), we provide the simulated distributions of impedance for electric realizations of (a), (c), (e), (g) and (i), respectively. The value of impedance at each node has been normalized to the maximum impedance.

# References

1. C. C. Adams. The knot book: An elementary introduction to mathematical theory of knots. (W. H. Freeman and Company. New York. 1994).
2. N. R. Cozzarelli, et. al., New Scientific Applications of Geometry and Topology, D. W. L. Sumners, Ed. (Proceedings of Symposia in Applied Mathematics, 1992).
3. R. H. Crowell, and R. H. Fox. Introduction to knot theory. (Springer-Verlag. New York. 1963).
4. L. H. Kauffman, and S. J Lomonaco Jr, Braiding operators are universal quantum gates. New J. Phys. 6, 134 (2004).
5. L. H. Kauffman, Topological quantum information, virtual Jones polynomials and Khovanov homology. New J. Phys. 13, 125007 (2013).
6. U. Tkalec, et. al., Reconfigurable Knots and Links in Chiral Nematic Colloids. Science. 333, 62 (2011).
7. S. Čopar, Topology and geometry of nematic braids. Phys. Rep. 538, 1 (2014).
8. H. K. Moffatt, R. L. Ricca, "Helicity and the Călugăreanu invariant" in Knots and Applications, L. H. Kauffman, Ed. (World Scientific, 1995).
9. J. B. Taylor, Relaxation of Toroidal Plasma and Generation of Reverse Magnetic Fields. Phys. Rev. Lett. 33, 1139 (1974).
10. D. Kleckner and W. T. Irvine, Creation and dynamics of knotted vortices. Nat. Phys. 9, 253 (2013).
11. D. Kleckner, L. H. Kauffman and W. T. Irvine, How superfluid vortex knots untie. Nat. Phys. 12, 650 (2016).
12. M. W. Scheeler, et. al., Complete measurement of helicity and its dynamics in vortex tubes. Science 357, 487 (2017).
13. V. P. Patil, et. al., Topological mechanics of knots and tangles. Science. 367, 71 (2020).
14. C. H. Lee, et. al., Imaging nodal knots in momentum space through topolectrical circuits. Nat. Commun. 11, 4385 (2020).
15. J. Leach, M. R. Dennis, J. Courtial, and M. J. Padgett, Vortex knots in light. New J. Phys. 7, 55 (2005).
16. W. T. M. Irvine, and D. Bouwmeester, Linked and knotted beams of light. Nat. Phys. 4, 716

(2008).

17. M. R. Dennis, et. al., Isolated optical vortex knots. Nat. Phys. 6, 118 (2010).

18. J. Hietarinta, J. Palmu, J. Jäykkä, and P. Pakkanen, Scattering of knotted vortices (Hopfions) in the Faddeev–Skyrme model. New J. Phys. 14, 013013 (2012).

19. F. Maucher, S. A. Gardiner, and I. G. Hughes, Excitation of knotted vortex lines in matter waves. New J. Phys. 18, 063016 (2016).

20. H. Larocque, et. al., Reconstructing the topology of optical polarization knots. Nat. Phys. 14, 1079 (2018).

21. E. Pisanty, et. al., Knotting fractional-order knots with the polarization state of light. Nat. Photonics 13, 569 (2019).

22. L. Wang, et. al., Ultrasmall Optical Vortex Knots Generated by Spin-Selective Metasurface Holograms. Adv. Optical Mater. 1900263, (2019).

23. H. Zhang, et. al., Creation of acoustic vortex knots. Nat. Commun. 11, 3956 (2020).

24. A. R. Klotz, B. W. Soh and P. S. Doyle, Motion of Knots in DNA Stretched by Elongational Fields. Phys. Rev. Lett. 120, 188003 (2018).

25. W. R. Taylor, A deeply knotted protein structure and how it might fold. Nature 406, 916 (2000).

26. Y. Arai, et. al., Tying a molecular knot with optical tweezers. Nature 399, 446 (1999).

27. N. Ponnuswamy, et. al., Discovery of an Organic Trefoil Knot. Science 338, 783 (2012).

28. D. S. Hall, et. al., Tying Quantum Knots. Nat. Phys. 12, 478 (2016).

29. J. J. Danon, et. al., Braiding a molecular knot with eight crossings. Science. 355, 159 (2017).

30. Y. Segawa, et. al., Topological molecular nanocarbons: all-benzene catenane and trefoil knot. Science 365, 272 (2019).

31. T. Ollikainen, et. al., Decay of a Quantum Knot. Phys. Rev. Lett. 123, 163003 (2019).

32. D. A. Leigh, et. al., Tying different knots in a molecular strand. Nature. 584, 562 (2020).

33. J. J. Champoux, DNA topoisomerases: Structure, function, and mechanism. Annu. Rev. Biochem. 70, 369 (2001).

34. V. V. Rybenkov, et. al., Simplification of DNA topology below equilibrium values by type II topoisomerases. Science. 277, 690 (1997).

35. Z. Liu, E. L. Zechiedrich and H. S. Chan, Inferring global topology from local juxtaposition

geometry: Interlinking polymer rings and ramifications for topoisomerase action. Biophys. J. 90, 2344 (2006).

36. Z. Liu, et. al., Topological information embodied in local juxtaposition geometry provides a statistical mechanical basis for unknotting by type-2 DNA topoisomerases. J. Mol. Biol. 361, 268 (2006).

37. Y. Burnier, et. al., Local selection rules that can determine specific pathways of DNA unknotting by type II DNA topoisomerases. Nucleic Acids Res. 35, 5223 (2007).

38. A. Barbensi, et. al., Grid diagrams as tools to investigate knot spaces and topoisomerase-mediated simplification of DNA topology. Sci. Adv. 6:eaay1458 (2020).

39. J. Ningyuan, et. al., Time- and Site-Resolved Dynamics in a Topological Circuit. Phys. Rev. X 5, 021031 (2015).

40. V. V. Albert, L. I. Glazman and L. Jiang, Topological Properties of Linear Circuit Lattices. Phys. Rev. Lett. 114, 173902 (2015).

41. S. Imhof, et. al., Topolectrical circuit realization of topological corner modes. Nat. Phys. 14, 925 (2018).

42. C. H. Lee, et. al., Topolectrical Circuits. Commun. Phys. 1, 39 (2018).

43. T. Hofmann, et. al., Phys. Rev. Lett. 122, 247702 (2019).

44. J. Bao, et. al., Topoelectrical circuit octupole insulator with topologically protected corner states. Phys. Rev. B 100, 201406 (2019).

45. N. A. Olekhno, et. al., Topological edge states of interacting photon pairs emulated in a topolectrical circuit. Nat. Commun. 11, 1436 (2020).

46. Y. Wang, et. al., Circuit Realization of a Four-Dimensional Topological Insulator. Nat. Commun. 11, 2356 (2020).

47. T. Helbig, et. al., Generalized bulk–boundary correspondence in non-Hermitian topolectrical circuits. Nat. Phys. 16, 747 (2020).

48. W. Zhang, et. al., Experimental Observation of Higher-Order Topological Anderson Insulators. Phys. Rev. Lett. 126, 146802 (2021).

49. P. Hoidn, R. B. Kusner, and A. Stasiak, Quantization of energy and writhe in self-repelling knots. New J. Phys. 4, 20 (2002).

50. C. Cerf, and A. Stasiak, Linear relations between writhe and minimal crossing number in

Conway families of ideal knots and links. New J. Phys. 5, 87 (2003).